\documentclass{ws-procs961x669}            
\usepackage{gensymb}

\newcommand{\rff}[1]{Fig.~\ref{#1}}

\begin{document}
\title{Magnetized black holes: the role of rotation, boost, and accretion in twisting the field lines and accelerating particles\footnote{Based on an invited contribution at 16th Marcel Grossmann Meeting, Session PT5 ``Dragging is never draggy: MAss and CHarge flows in GR'' (id. \#393), 5--10 July 2021.}}

\author{Ond\v{r}ej Kop\'{a}\v{c}ek and Vladim\'{\i}r Karas}

\address{Astronomical Institute, Czech Academy of Sciences,\\
Bo\v{c}n\'{i}~II~1401, CZ-141\,00~Prague, Czech~Republic\\
E-mail: kopacek@ig.cas.cz, vladimir.karas@asu.cas.cz\\
http://astro.cas.cz
}

\begin{abstract}
Combined influence of rotation of a black hole and ambient magnetic fields creates conditions for powerful astrophysical processes of accretion and outflow of matter which are observed in many systems across the range of masses; from stellar-mass black holes in binary systems to supermassive black holes in active galactic nuclei. We study a simplified model of outflow of electrically charged particles from the inner region of an accretion disk around a spinning (Kerr) black hole immersed in a large-scale magnetic field. In particular, we consider a non-axisymmetric magnetosphere where the field is misaligned with the rotation axis. In this contribution we extend our previous analysis of acceleration of jet-like trajectories of particles escaping from bound circular orbits around a black hole. While we have previously assumed the initial setup of prograde (co-rotating) orbits, here we relax this assumption and we also consider retrograde (counter-rotating) motion. We show that the effect of {\em counter-rotation may considerably increase the probability of escape from the system, and it allows more efficient acceleration of escaping particles to slightly higher energies compared to the co-rotating disk}.
\end{abstract}


\bodymatter

\section{Introduction}\label{intro}
In this contribution we further study astrophysically relevant class of escaping (jet-like accelerated) trajectories in a dynamical system of a magnetized rotating black hole. Electrically charged particles following such orbits may escape the attraction of the black hole in a collimated outflow and they may be accelerated to very high energies. In Paper I\cite{kopacek18} we adopted a simple axisymmetric model consisting of a rotating black hole in an asymptotically uniform magnetic field aligned with the rotation axis. Initially neutral particles are supposed to follow circular orbits in the equatorial plane (and freely fall below ISCO). At a particular ionization radius, these particles obtain electric charge and their dynamics changes due to the presence of electromagnetic field, and escape from the accretion disk along high-velocity jet-like trajectories becomes possible. However, the efficiency of the acceleration process remains only moderate in the axisymmetric setup.\cite{kopacek18}

In Paper II\cite{kopacek20} we relax the assumption of the perfect spin-magnetic alignment and consider a generalized non-axisymmetric model. The effect of non-axisymmetry in electro-vacuum magnetospheres of compact objects is known to introduce dynamically relevant features like magnetic null-points,\cite{kopacek18b} and breaking the axial symmetry also appears to strongly affect the acceleration process in the current model. In particular, the number of escaping orbits generally increases and acceleration to ultrarelativistic energies becomes possible in the oblique magnetosphere.\cite{kopacek20, karas21} Furthermore,\cite{2012CQGra..29c5010K} by employing a boost transformation of the system into a frame in translation motion, we explored the electromagnetic aspects of the linear (kick) velocity of the black hole.  Refer to enlightening discussions by various authors\cite{karas13,2018arXiv181008066A} and see also the relevant context of these solutions.\citep{2020PhRvD.101b8501G}

In the above-mentioned previous papers, we numerically studied escaping orbits and we assessed in detail the role of a broad variety of parameters of the model. Nevertheless, the assumption that initially neutral matter co-rotates with the central black hole, was adopted in all previous analyses. Here we relax this assumption and we also consider counter-rotating (retrograde) circular orbits as an initial setup of neutral particles. We compare the both cases (of co-rotating vs.\ counter-rotating) in terms of the efficiency of the formation and acceleration of the individual trajectories and the emerging outflow.

The role of counter-rotation in processes of accretion and outflow near compact objects has been only partially explored in the literature and various relevant aspects of accretion of counter-rotating matter has been studied. In particular, high-resolution hydrodynamic simulations suggest that counter-rotating structure may largely increase the accretion rate.\cite{dyda15} Moreover, general relativistic magnetohydrodynamic (GRMHD) simulations show that accretion of counter-rotating matter may produce conditions for highly efficient jet launching.\cite{koide00} We also note that counter-rotating high-velocity outflows are indeed reported in various systems, such as the spiral galaxy NGC 1068.\cite{imp19}

The paper is organized as follows. In Section~\ref{spec} we specify the employed model of oblique black hole magnetosphere, we review the equations of motion and describe the charging process of initially neutral matter. Jet-like trajectories of charged particles escaping from counter-rotating orbits are discussed in Section~\ref{escape} with the focus on the direct comparison with the co-rotating analogues. Results are summarized and briefly discussed in Section~\ref{concl}. As the initial setup of the model investigated in the present paper is analogous to that of Papers~I and II, we refer to the more complete introduction and references presented therein.

\section{Model specification}\label{spec}
We consider a spacetime around a rotating, axially-symmetric, uncharged black hole of mass $M$ and spin $a$ described by the Kerr metric, which may be expressed in Boyer--Lindquist coordinates $x^{\mu}= (t,\:r, \:\theta,\:\varphi)$ as follows:\cite{mtw,kerr63}
\begin{equation}
\label{metric}
ds^2=-\frac{\Delta}{\Sigma}\:[dt-a\sin{\theta}\,d\varphi]^2+\frac{\sin^2{\theta}}{\Sigma}\:[(r^2+a^2)d\varphi-a\,dt]^2+\frac{\Sigma}{\Delta}\;dr^2+\Sigma d\theta^2,
\end{equation}
where $\Delta(r)\equiv r^2-2Mr+a^2$ and $\Sigma(r,\theta)\equiv r^2+a^2\cos^2\theta$. Roots of $\Delta(r)$ locate the outer (+) and the inner (-) horizons as: $r_{\pm}=M\pm\sqrt{M^2-a^2}$. Geometrized units are used throughout the paper; values of basic constants (gravitational constant $G$, speed of light $c$, Boltzmann constant $k$, and Coulomb constant $k_C$) therefore equal unity, $G=c=k=k_C=1$.

The black hole is supposed to be weakly magnetized by an external, asymptotically uniform magnetic field with an asymptotic strength $B$ which has an arbitrary inclination (angle $\alpha$) with respect to the rotation axis. Vector potential $A_{\mu}=(A_t,A_r,A_{\theta},A_{\varphi})$ of a corresponding test-field solution is given as:\cite{bicak85}

\begin{eqnarray}
\label{empot1}
A_t&=&\frac{B_{z}aMr}{\Sigma}\left(1+\cos^2\theta\right)-B_{z}a+\frac{B_xaM\sin\theta\cos\theta}{\Sigma}\left(r\cos\psi-a\sin\psi\right),\\
A_r&=&-B_x(r-M)\cos\theta\sin\theta\sin\psi,\label{empot2}\\
A_{\theta}&=&-B_xa(r\sin^2\theta+M\cos^2\theta)\cos\psi-B_x(r^2\cos^2\theta\\
\nonumber& &-Mr\cos2\theta+a^2\cos2\theta)\sin\psi,\label{empot3}\\
A_{\varphi}&=&B_z\sin^2\theta\left[\frac{1}{2}(r^2+a^2)-\frac{a^2Mr}{\Sigma}(1+\cos^2\theta)\right]\label{empot4}\\
\nonumber& &-B_x\sin\theta\cos\theta\left[\Delta\cos\psi+\frac{(r^2+a^2)M}{\Sigma}\left(r\cos\psi-a\sin\psi\right)\right],
\end{eqnarray}
where $B_z$ denotes the component parallel to the rotation axis, while $B_x$ corresponds to the perpendicular component, i.e., $B_z=B\cos{\alpha}$ and $B_x=B\sin{\alpha}$. Setting $B_x=0$ reduces the above vector potential $A_{\mu}$ to the axisymmetric solution \cite{wald74} employed in Paper~I. Azimuthal coordinate $\psi$ of Kerr ingoing coordinates is expressed in Boyer--Lindquist coordinates as follows:
\begin{equation}
\label{kicpsi}
\psi=\varphi+\frac{a}{r_{+}-r_{-}}\ln{\frac{r-r_{+}}{r-r_{-}}}.
\end{equation}

The Hamiltonian $\mathcal{H}$ of a particle of electric charge $q$ and rest mass $m$ in the field $A_{\mu}$ and metric $g^{\mu\nu}$ is defined as:\cite{mtw}
\begin{equation}
\label{hamiltonian}
\mathcal{H}=\textstyle{\frac{1}{2}}g^{\mu\nu}(\pi_{\mu}-qA_{\mu})(\pi_{\nu}-qA_{\nu}),
\end{equation}
where $\pi_{\mu}$ is the generalized (canonical) momentum. The equations of motion are expressed as:
\begin{equation}
\label{hameq}
\frac{{\rm d}x^{\mu}}{{\rm d}\lambda}\equiv p^{\mu}=
\frac{\partial \mathcal{H}}{\partial \pi_{\mu}},
\quad 
\frac{d\pi_{\mu}}{d\lambda}=-\frac{\partial\mathcal{H}}{\partial x^{\mu}},
\end{equation}
where $\lambda\equiv\tau/m$ is dimensionless affine parameter ($\tau$ denotes the
proper time). Employing the first equation we obtain the kinematical four-momentum as: $p^{\mu}=\pi^{\mu}-qA^{\mu}$, and the conserved value of the Hamiltonian is therefore given as: $\mathcal{H}=-m^2/2$. System is stationary and the time component of canonical momentum $\pi_t$ is therefore an integral of motion which equals (negatively taken) energy of the test particle $\pi_t\equiv-E$. In the rest of paper we switch to specific quantities $E/m\rightarrow E$, $q/m\rightarrow q$ which corresponds to setting the rest mass of the particle $m=1$ in the formulas.

\begin{figure}[ht]
\center
\includegraphics[scale=.38]{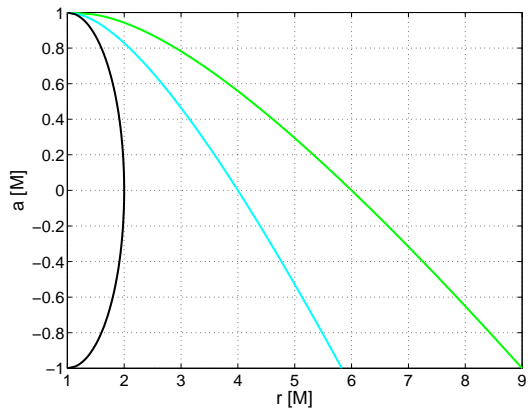}
\caption{Position of the black hole's outer horizon (black), marginally bound circular orbit (cyan) and ISCO (green). Negative spin values correspond to the counter-rotating orbits.}
\label{isco}
\end{figure}

As an initial configuration of electrically neutral particles, we consider circular Keplerian orbits specified by the values of constants of motion, i.e., by specific energy $E_{\rm Kep}$ and specific angular momentum $L_{\rm Kep}$ given as follows:\cite{bardeen72}

\begin{equation}
 \label{kepconst}
E_{\rm Kep}=\frac{r^2-2Mr\pm a \sqrt{Mr}}{r\sqrt{r^2-3Mr\pm 2a\sqrt{Mr}}},\;\;\;\;L_{\rm Kep}=\frac{\pm\sqrt{M} (r^2+a^2\mp 2a\sqrt{Mr})}{\sqrt{r(r^2-3Mr\pm 2a\sqrt{Mr})}},
\end{equation}
where the upper signs are valid for the prograde (co-rotating) orbits and the lower ones for the retrograde (counter-rotating) orbits. 
Circular geodesics remain stable only above the innermost stable circular orbit (ISCO):
\begin{equation}
 \label{rms}
r_{\rm{ISCO}}=M\left(3+Z_2\mp\sqrt{(3-Z_1)(3+Z_1+2Z_2)}\right),
\end{equation}
where $Z_1\equiv1+\left(1-\frac{a^2}{M^2}\right)^{1/3}\left[\left(1+\frac{a}{M}\right)^{1/3}+\left(1-\frac{a}{M}\right)^{1/3}\right]$ and $Z_{2} \equiv \sqrt{\frac{3a^2}{M^2}+Z_1^2}$. Positions of the ISCO and marginally bound circular orbit\cite{bardeen72} are plotted in \rff{isco}.

Below ISCO the geodesics are supposed to turn into freely falling inspirals maintaining the energy and angular momentum corresponding to the ISCO radius, i.e., the particles keep $E=E_{\rm Kep}(r_{\rm ISCO})$ and $L=L_{\rm Kep}(r_{\rm ISCO})$ during their infall while the radial velocity $u^r$ is obtained from the normalization condition $u^{\mu}u_{\mu}=-1$.

In our model, we suppose that initially neutral elements undergo a sudden charging process (e.g.,  due to photoionization) at a given radius $r_0$ and obtain specific electric charge $q$. While the change of the rest mass $m$ is considered as negligible, the particle dynamics changes due to the presence of the electromagnetic field.\cite{stuchlik16} In particular, the conserved value of energy is modified as:
\begin{equation}
\label{newenergy}
E=E_{\rm Kep}-qA_t,
\end{equation}
while the values of spatial components of canonical momentum are changed as:
\begin{equation}
\label{newmomenta}
\pi_r=\pi_r^0+qA_r,\;\;\;\pi_{\theta}=qA_{\theta},\;\;\;\pi_{\varphi}=L_{\rm Kep}+qA_{\varphi}, 
\end{equation}
where $\pi_r^0$ is zero for particles ionized above/at ISCO and for infalling particles with ionization radius $r_0 < r_{\rm{ISCO}}$ the value is calculated from the normalization condition.

\begin{figure}[ht]
\center
\includegraphics[scale=.28]{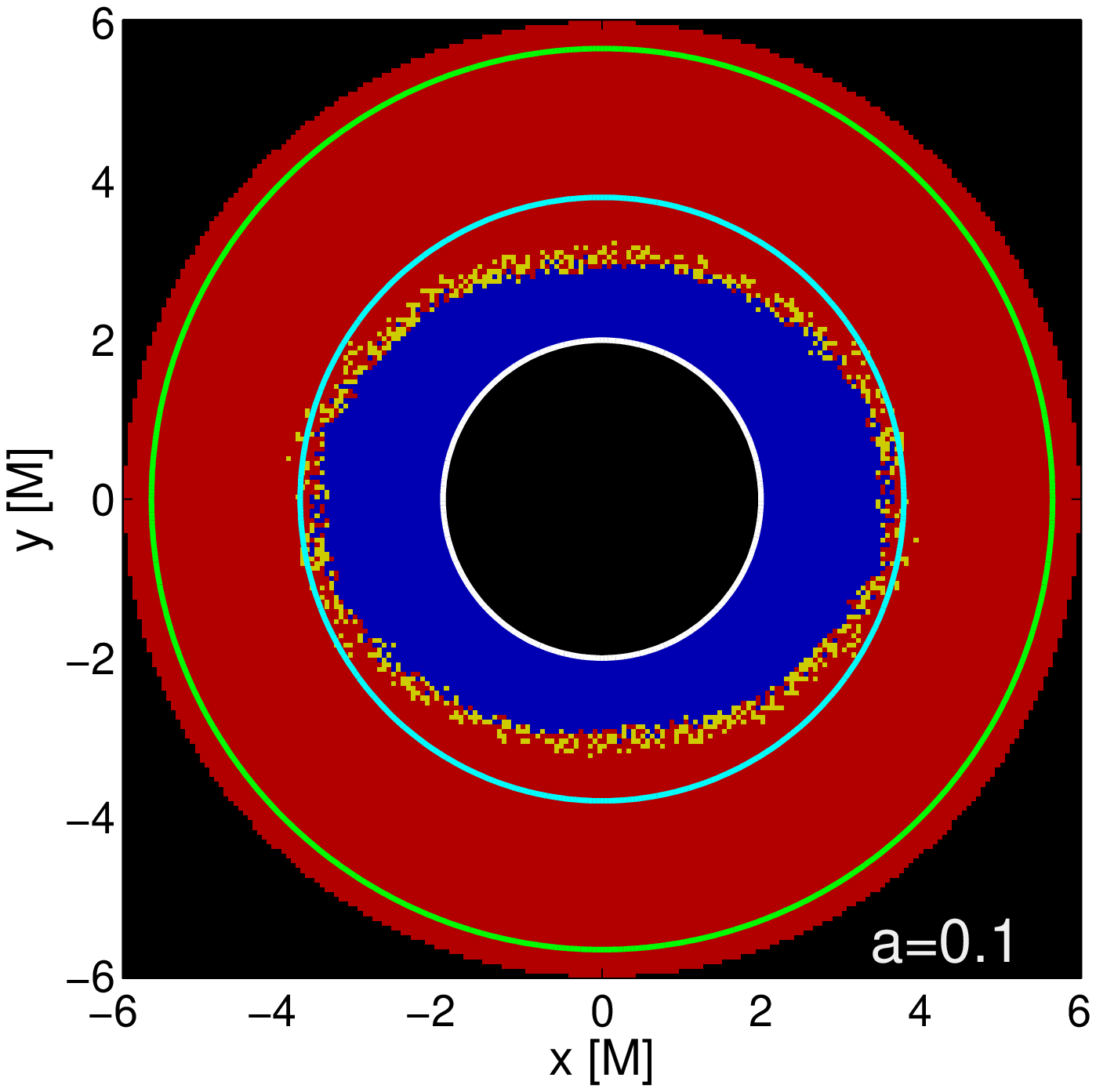}
\includegraphics[scale=.28]{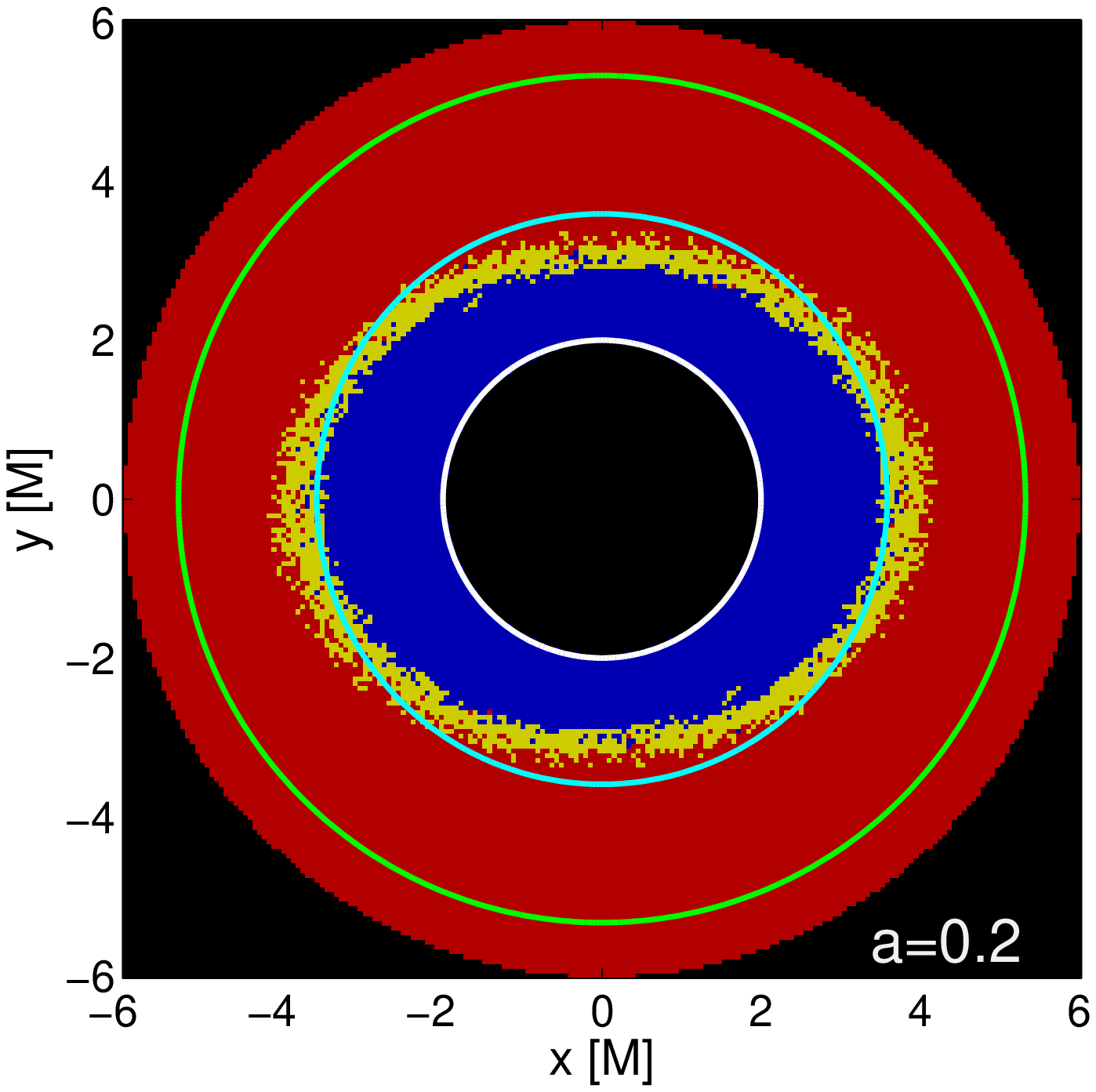}
\includegraphics[scale=.28]{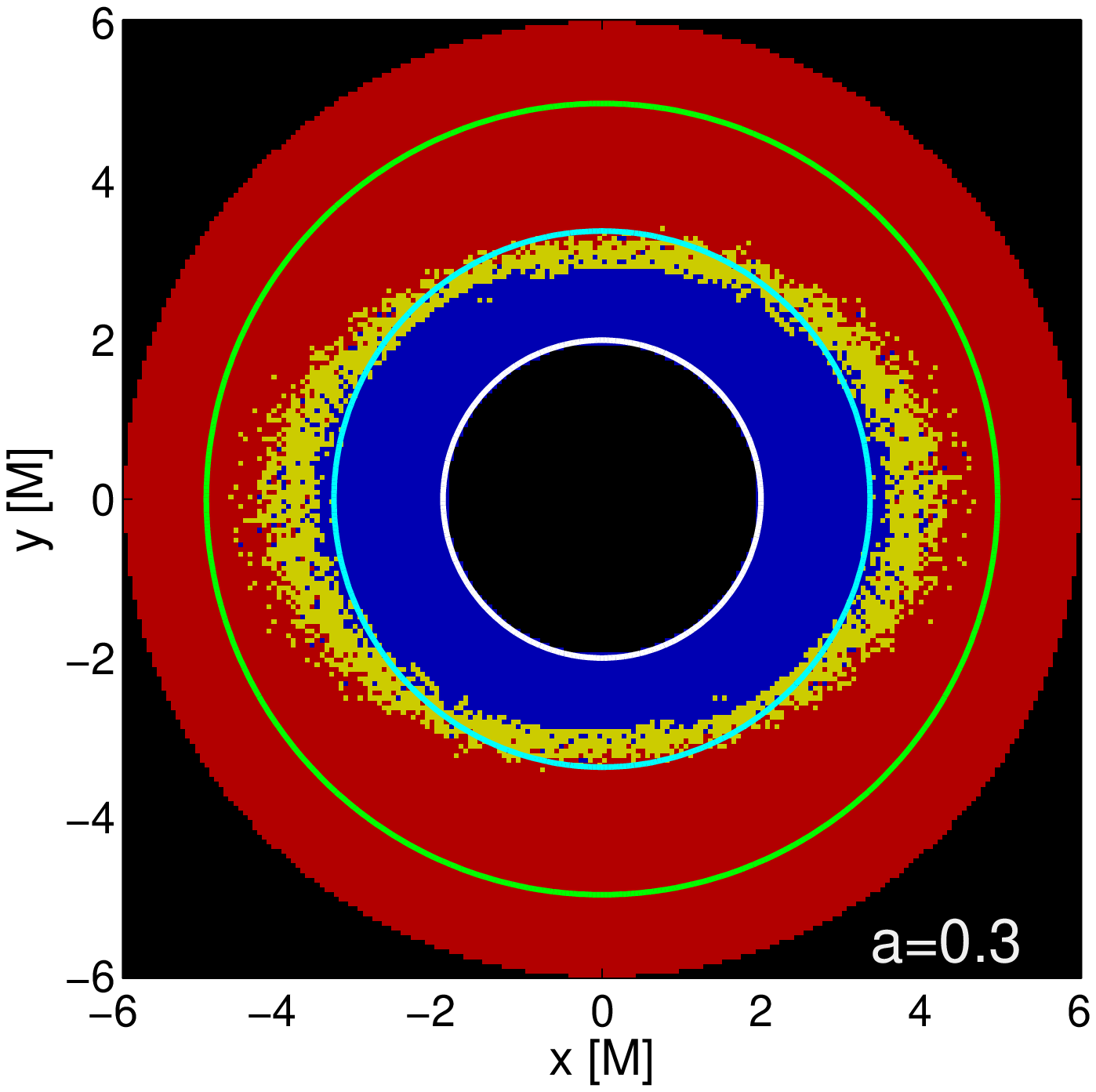}\\
\includegraphics[scale=.28]{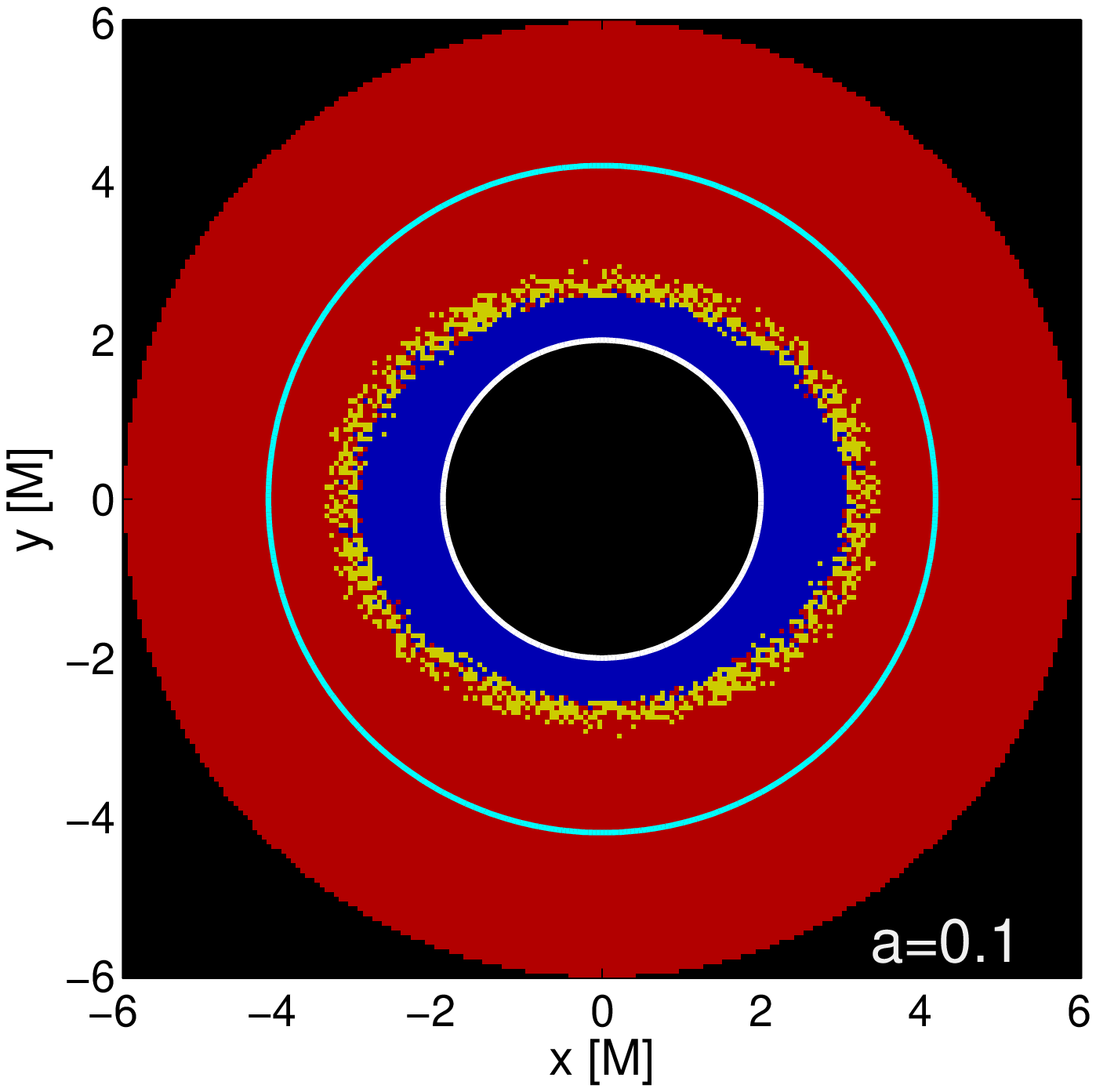}
\includegraphics[scale=.28]{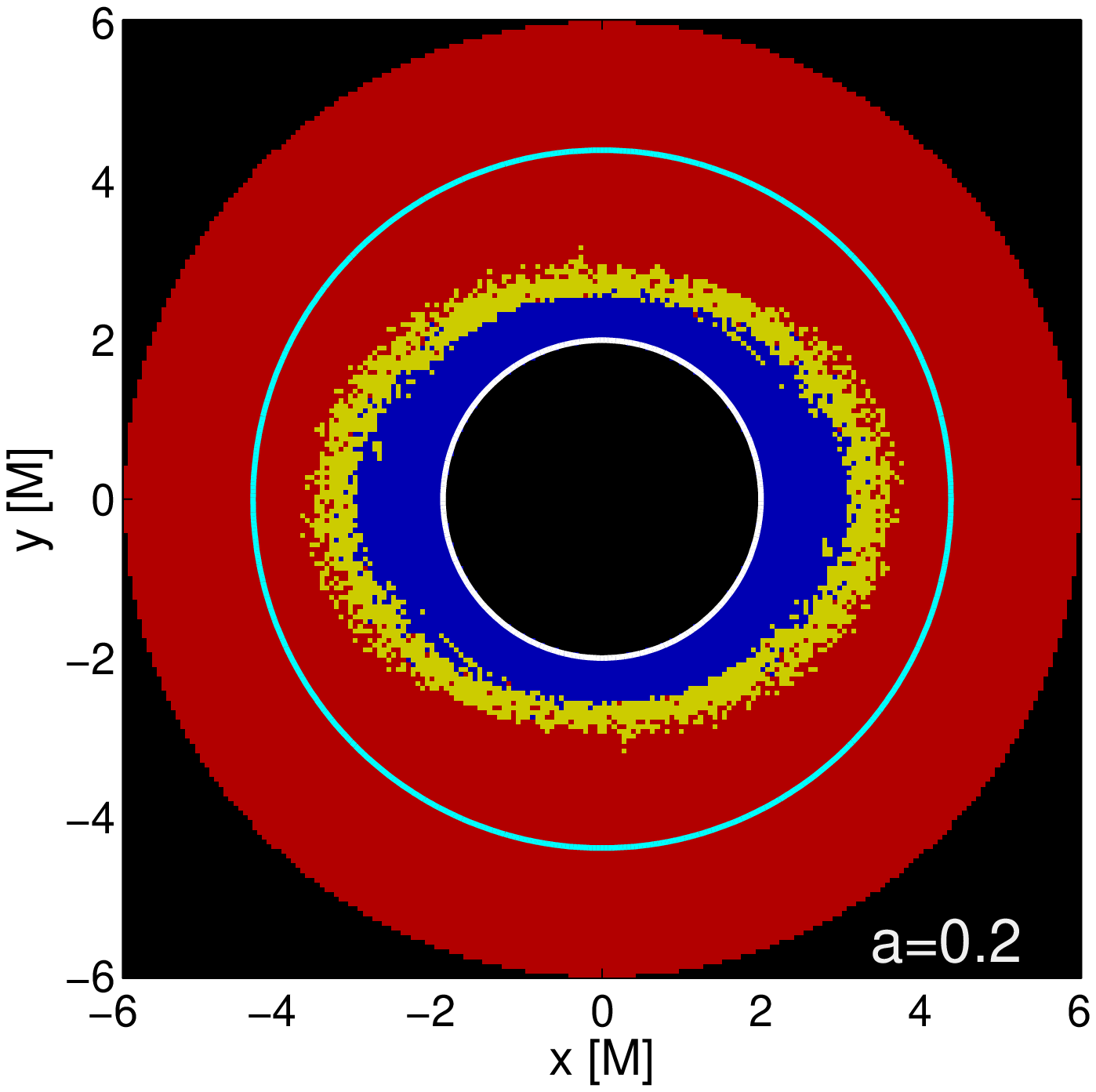}
\includegraphics[scale=.28]{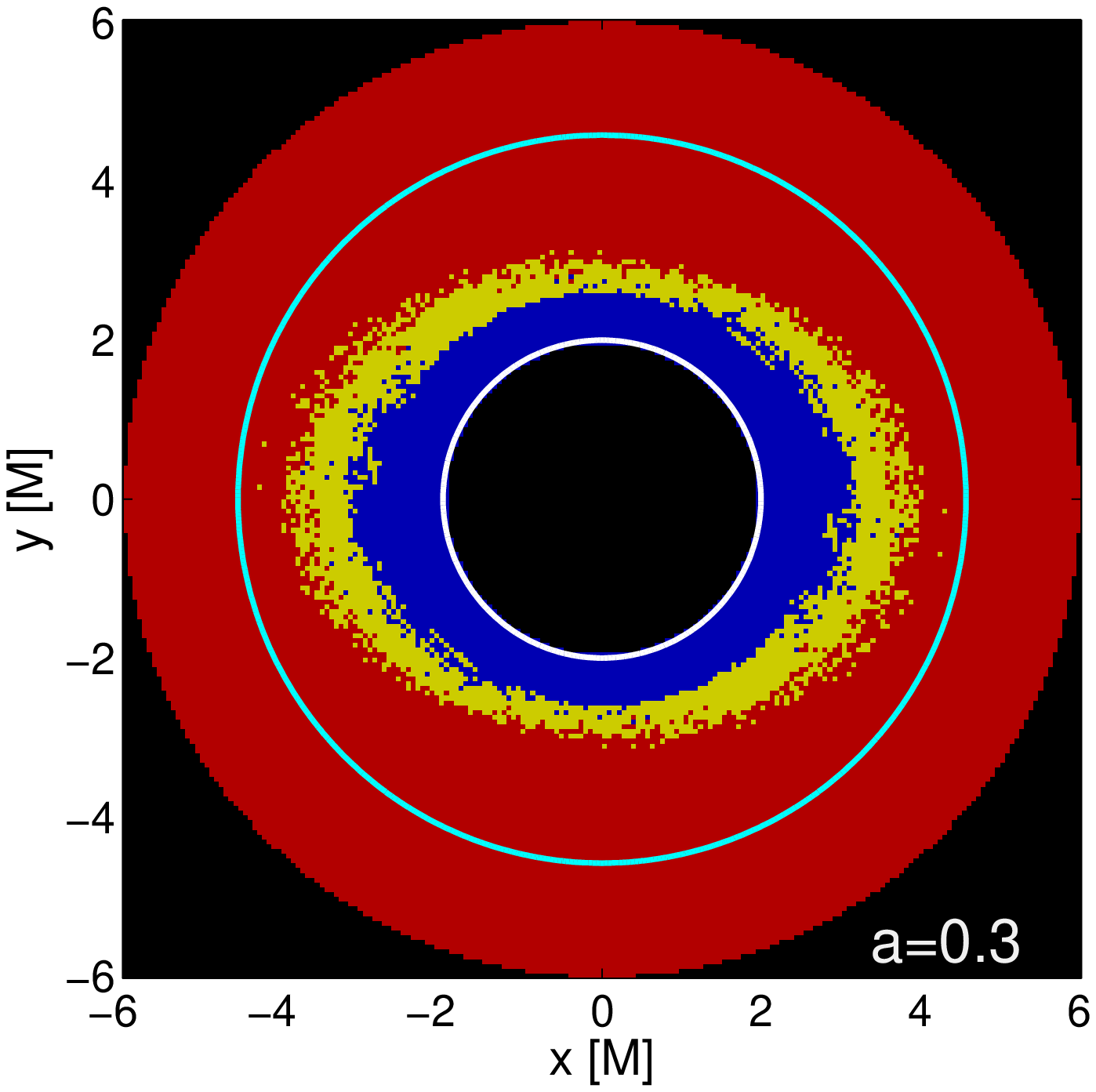}
\caption{Comparison of initially co-rotating orbits (upper row) with counter-rotating orbits (lower row). Color-coding: blue for plunging orbits, red for stable ones, and yellow for escaping trajectories. The green circle denotes the ISCO, the cyan circle marks the radius of the marginally bound orbit and the white one shows the ergosphere. The inner black region marks the horizon of the black hole. The parameters of the system are $\alpha=35\degree$ and $qB=-5\,M^{-1}$. The magnetic field is inclined in the positive $x$-direction.}
\label{azi1}
\end{figure}

\section{Escape zones}\label{escape}
The dynamics of initially neutral particles on stable circular orbits (above ISCO) or on a plunging trajectory (below ISCO) changes abruptly as they obtain the electric charge and become affected by the electromagnetic field given by Eqs. (\ref{empot1})-(\ref{empot4}). As a result, some stable particles may become plunging, while some plunging trajectories are stabilized by the field. Moreover, it appears that particles may also become unbound and escape to infinity in jet-like trajectories.

We have previously analyzed escaping particles in this model and, in particular, in Paper I\cite{kopacek18} we considered the axisymmetric system ($B_x=0$), while in the Paper II\cite{kopacek20} the general non-axisymmetric configuration was studied. We have used the method of effective potential to find the necessary conditions for the escape of particles. It appears that for the parallel orientation of the spin axis $z$ and magnetic field component $B_z$, only negatively charged particles may escape (while positively charged escape for the anti-parallel orientation). Non-zero spin $a$ and $B_z$ are required for the escape and final Lorentz factor $\gamma$ is found to be an increasing function of the spin, specific charge $|q|$ and $|B_z|$ while it decreases with the ionization radius $r_0$. 

\begin{figure}[ht]
\center
\includegraphics[scale=.28]{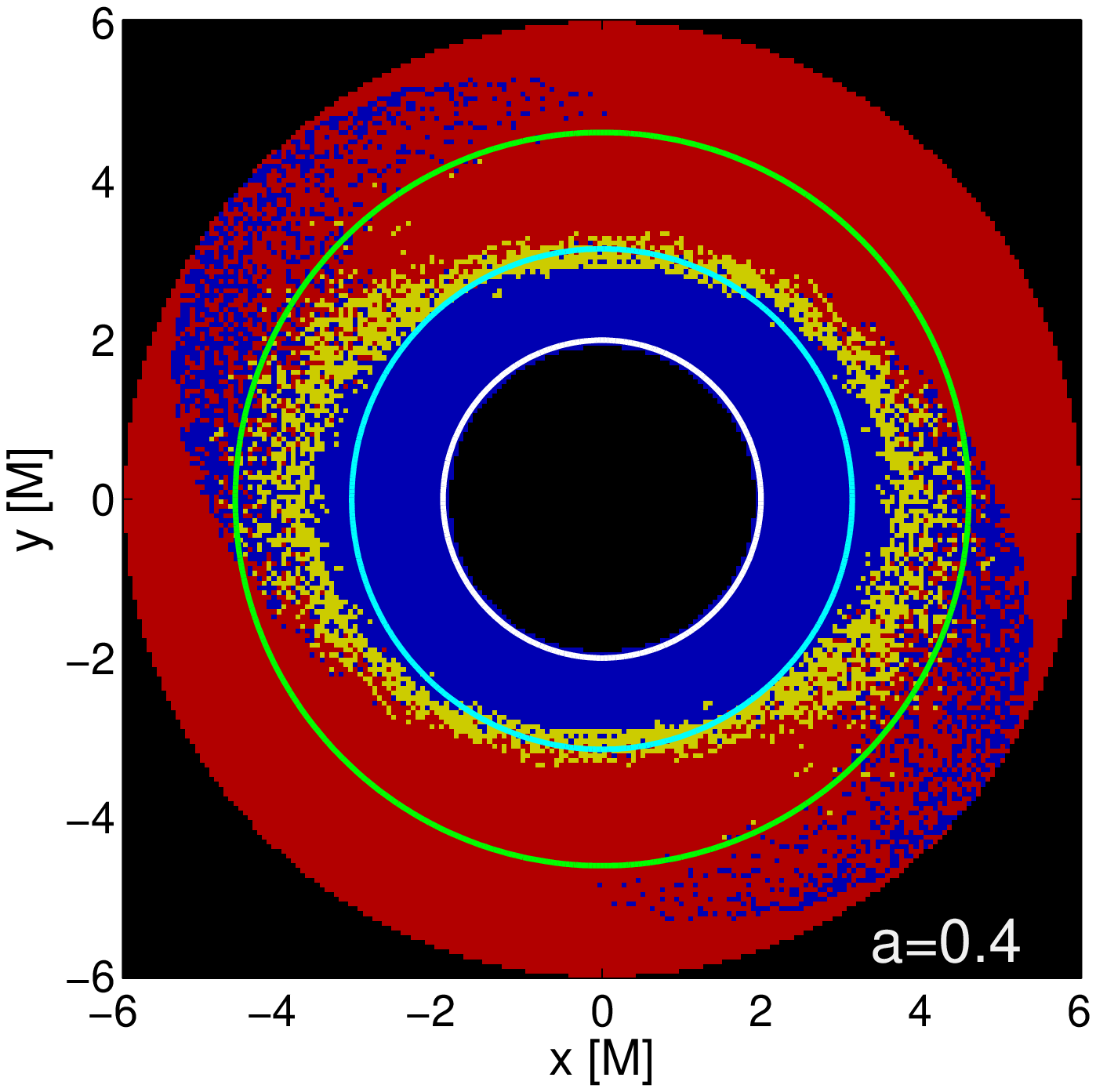}
\includegraphics[scale=.28]{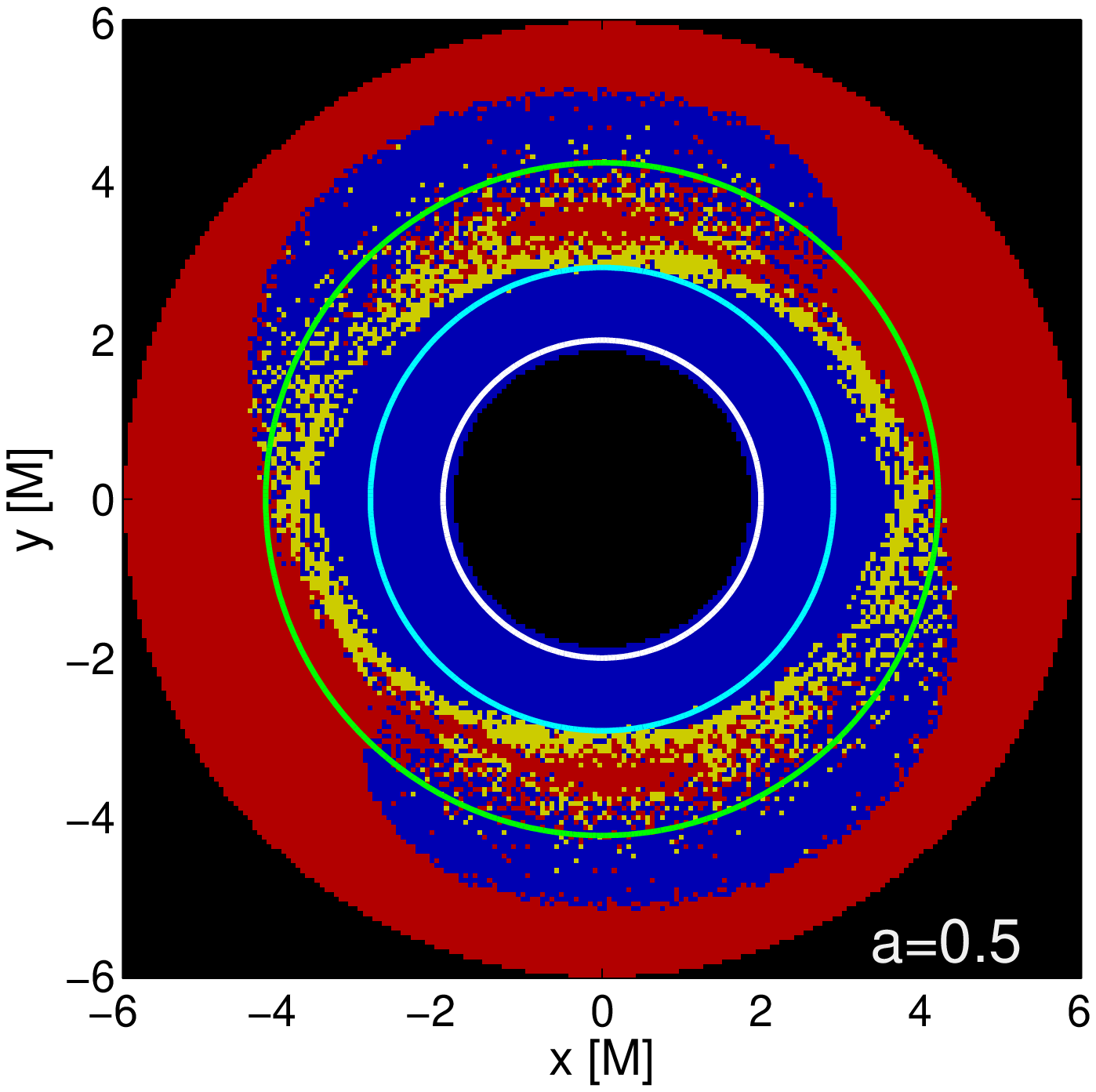}
\includegraphics[scale=.28]{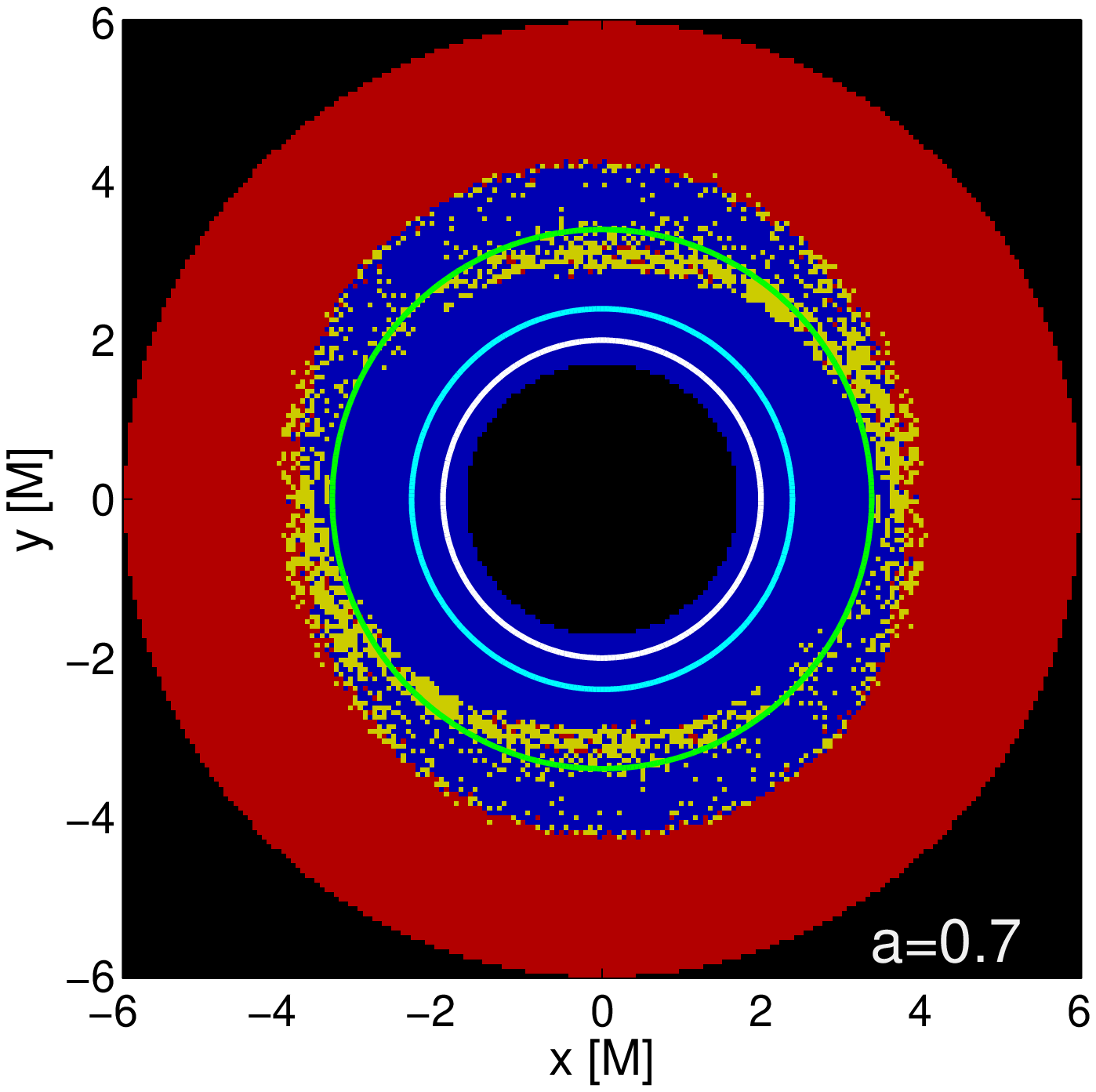}\\
\includegraphics[scale=.28]{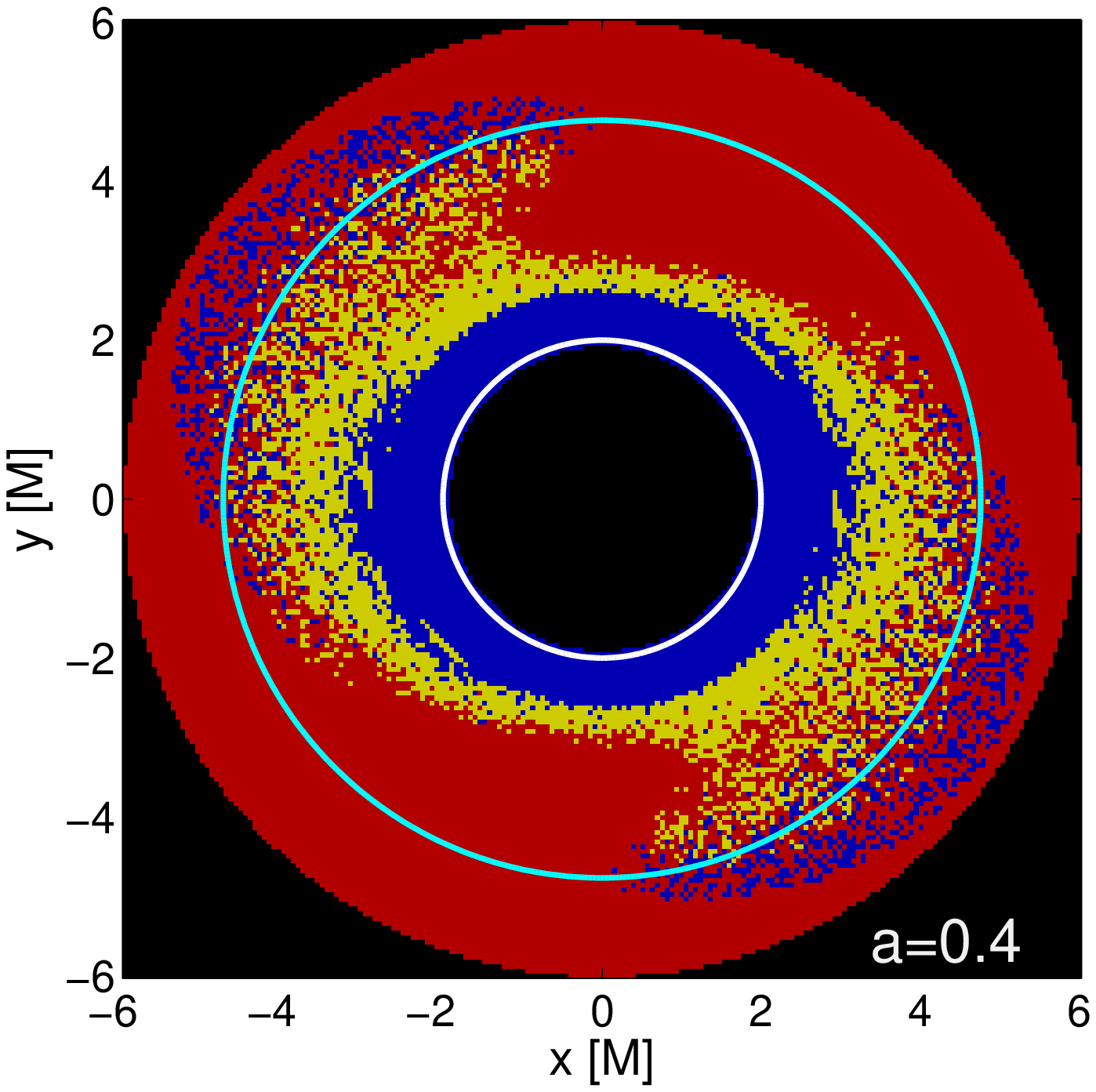}
\includegraphics[scale=.28]{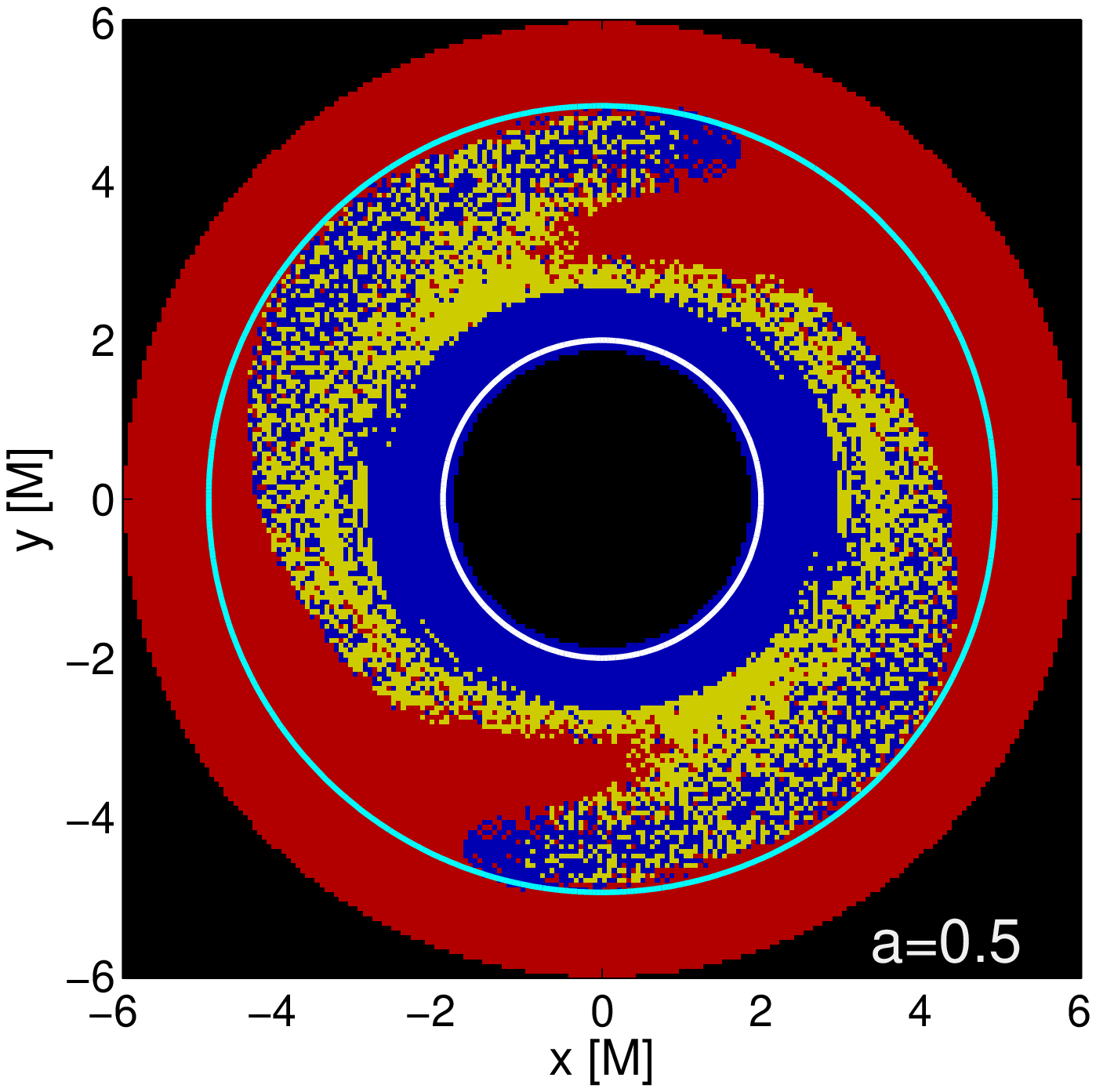}
\includegraphics[scale=.28]{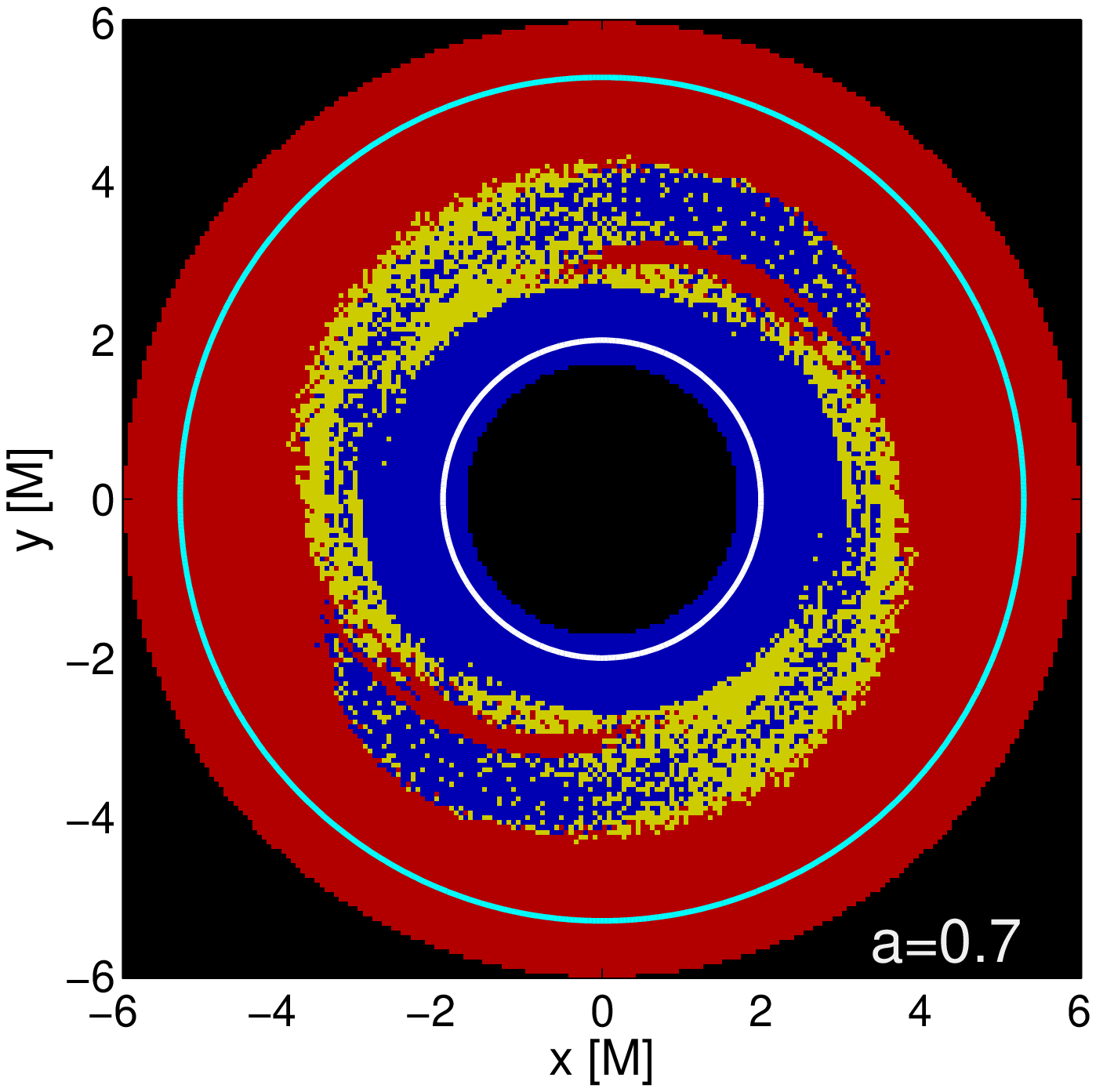}
\caption{Comparison of initially co-rotating orbits (upper row) with counter-rotating orbits (lower row). The color-coding and parameters as in Fig.~\ref{azi1}.}
\label{azi2}
\end{figure}

However, as the above conditions are necessary but not sufficient, we need to investigate the dynamic system numerically in order to determine which particles actually follow the escaping trajectories. Initial locations of escaping particles form {\em escape zones} whose emergence and evolution with respect to the parameters of the system was discussed in our previous papers.\cite{karas21,kopacek18,kopacek20} Efficiency of the acceleration mechanism was quantified by the final Lorentz factor of escaping particles leading to the conclusion that also ultrarelativistic velocities with very high energies may be achieved within the non-axisymmetric model. 

In this contribution, we extend the previous discussion and consider a modified setup. In particular, here we suppose that the initially neutral matter may follow counter-rotating geodesics, i.e., unlike the previous papers, here we also consider the lower signs in Eqs. (\ref{kepconst}) and (\ref{rms}). The ISCO of counter-rotating orbits is located generally farther from the horizon compared to the co-rotating ones (see Fig.~\ref{isco}). In particular, no stable retrograde circular orbits around rotating black hole are allowed for $r\leq 6\,M$ and the ISCO location grows with the spin up to $r_{\rm ISCO}= 9\,M$ for the maximally rotating black hole with $a=M$. 

We assume that below ISCO the initially neutral particle falls freely towards the horizon keeping the energy and angular momentum corresponding to the ISCO radius, while the radial velocity is computed from the normalization condition. At the ionization radius $r_0$ the particle obtains the electric charge and its dynamics changes due to magnetic field. Three main scenarios may occur then; the particle plunges into the horizon, follows the stable orbit around the black hole or escapes to infinity. 

\begin{figure}[ht]
\center
\includegraphics[scale=.28]{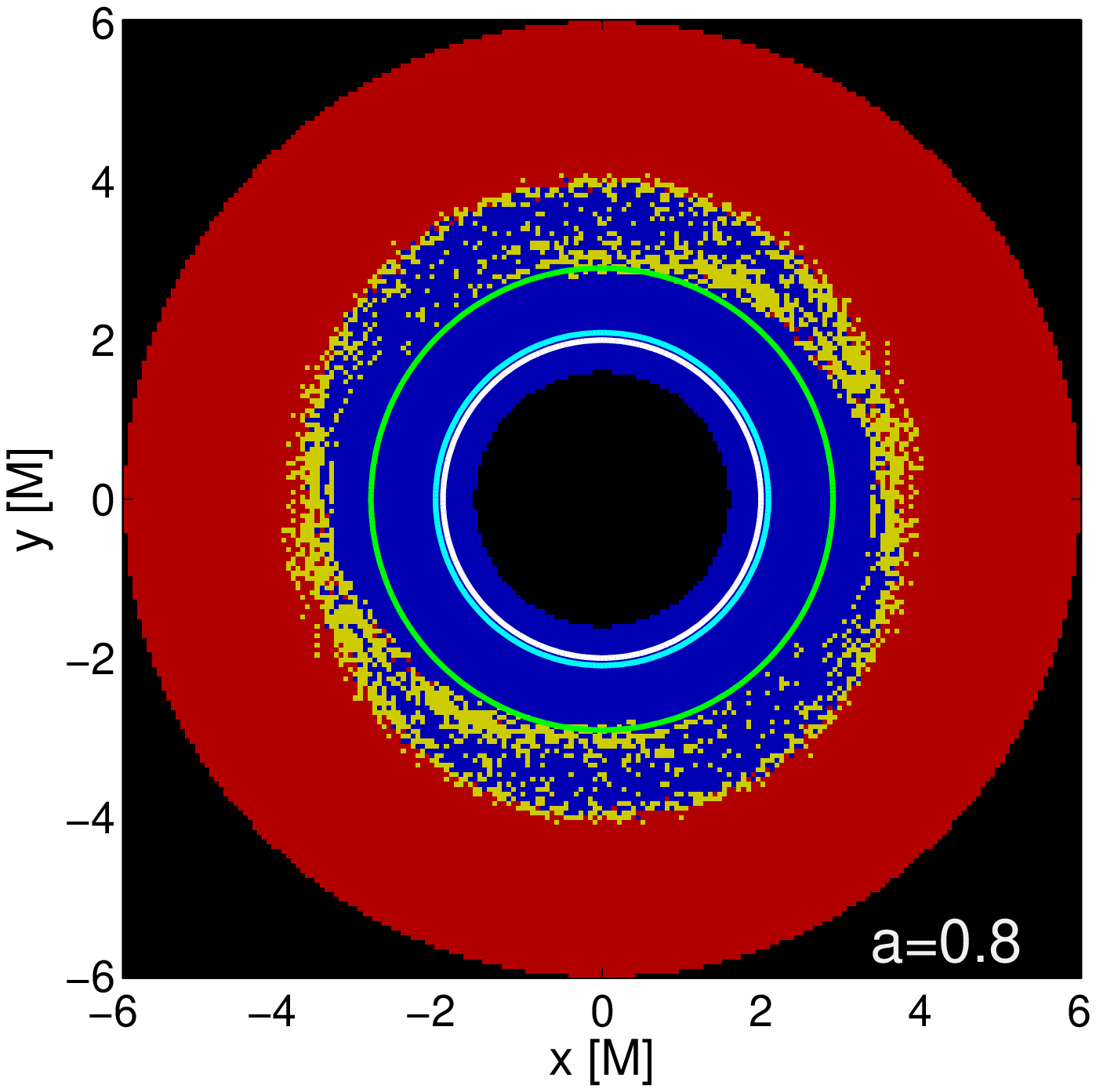}
\includegraphics[scale=.28]{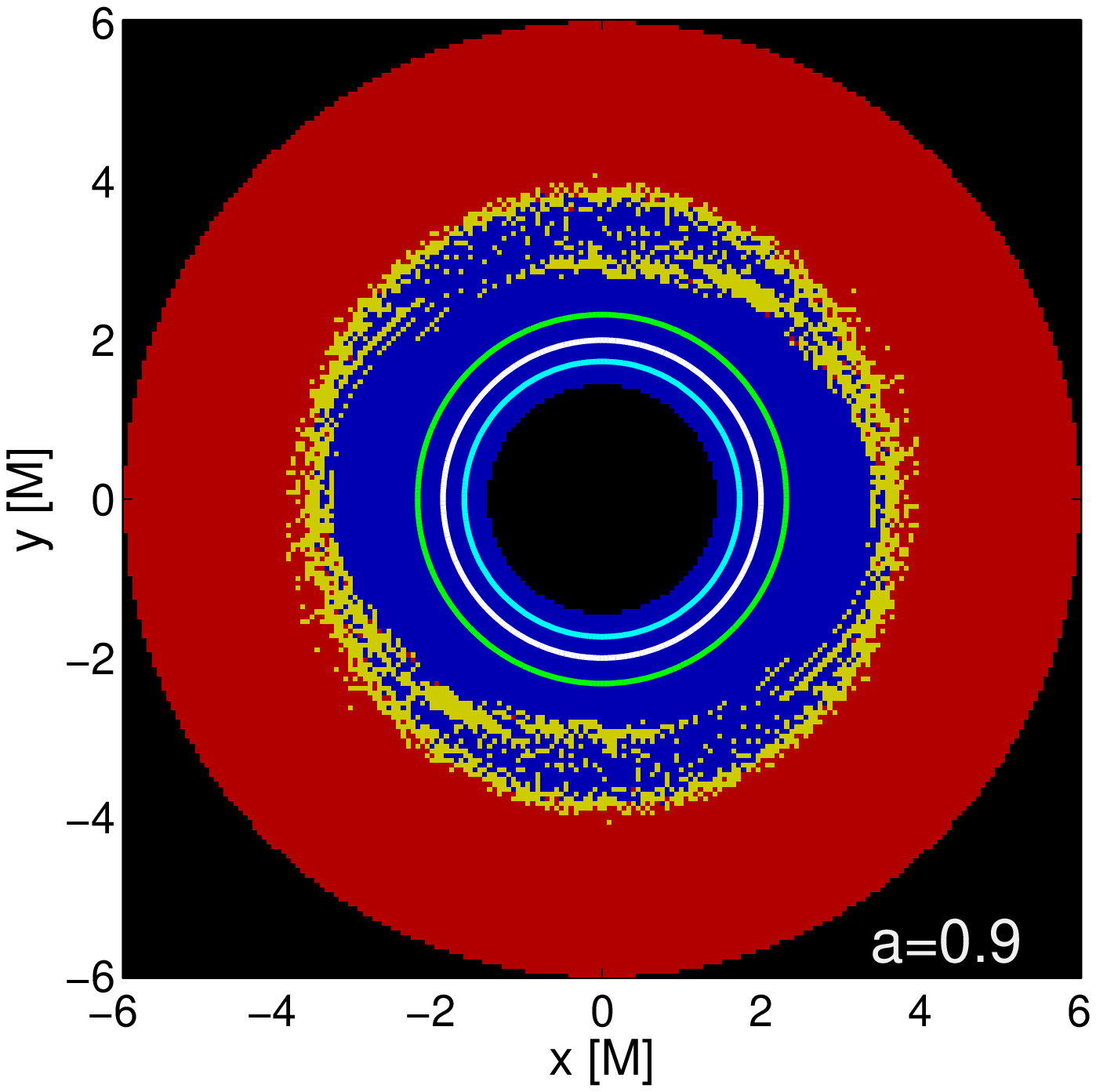}
\includegraphics[scale=.28]{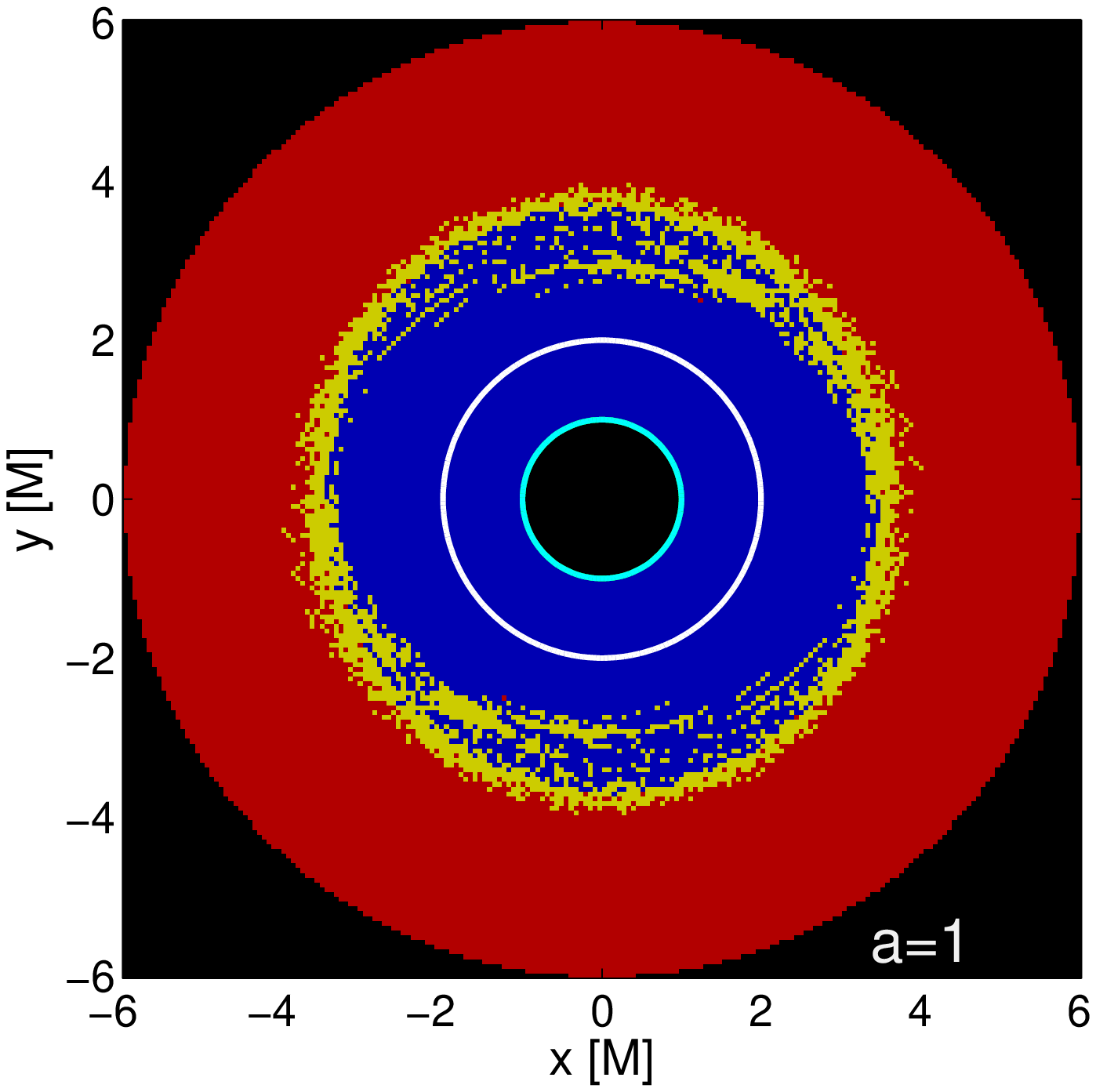}\\
\includegraphics[scale=.28]{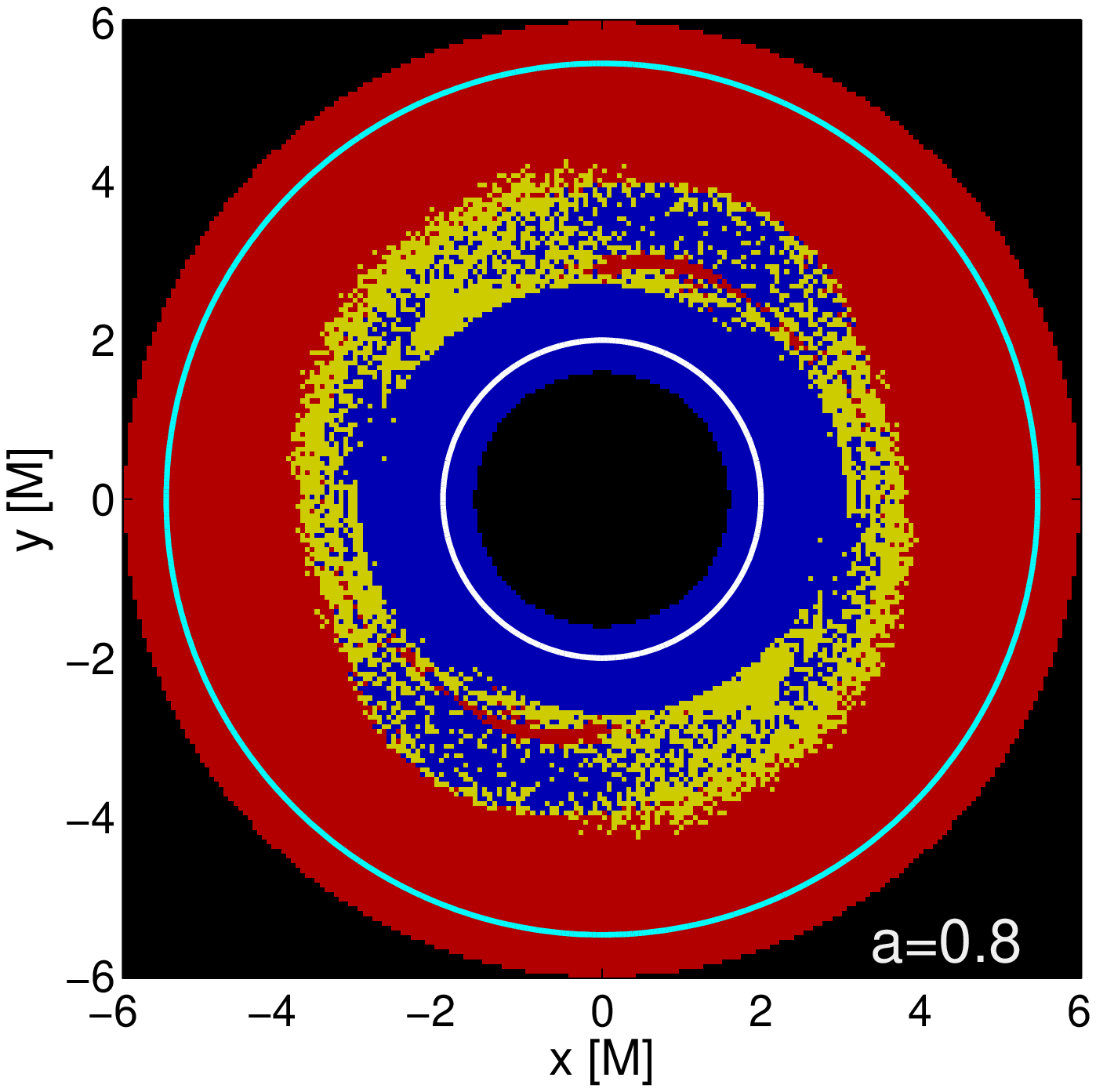}
\includegraphics[scale=.28]{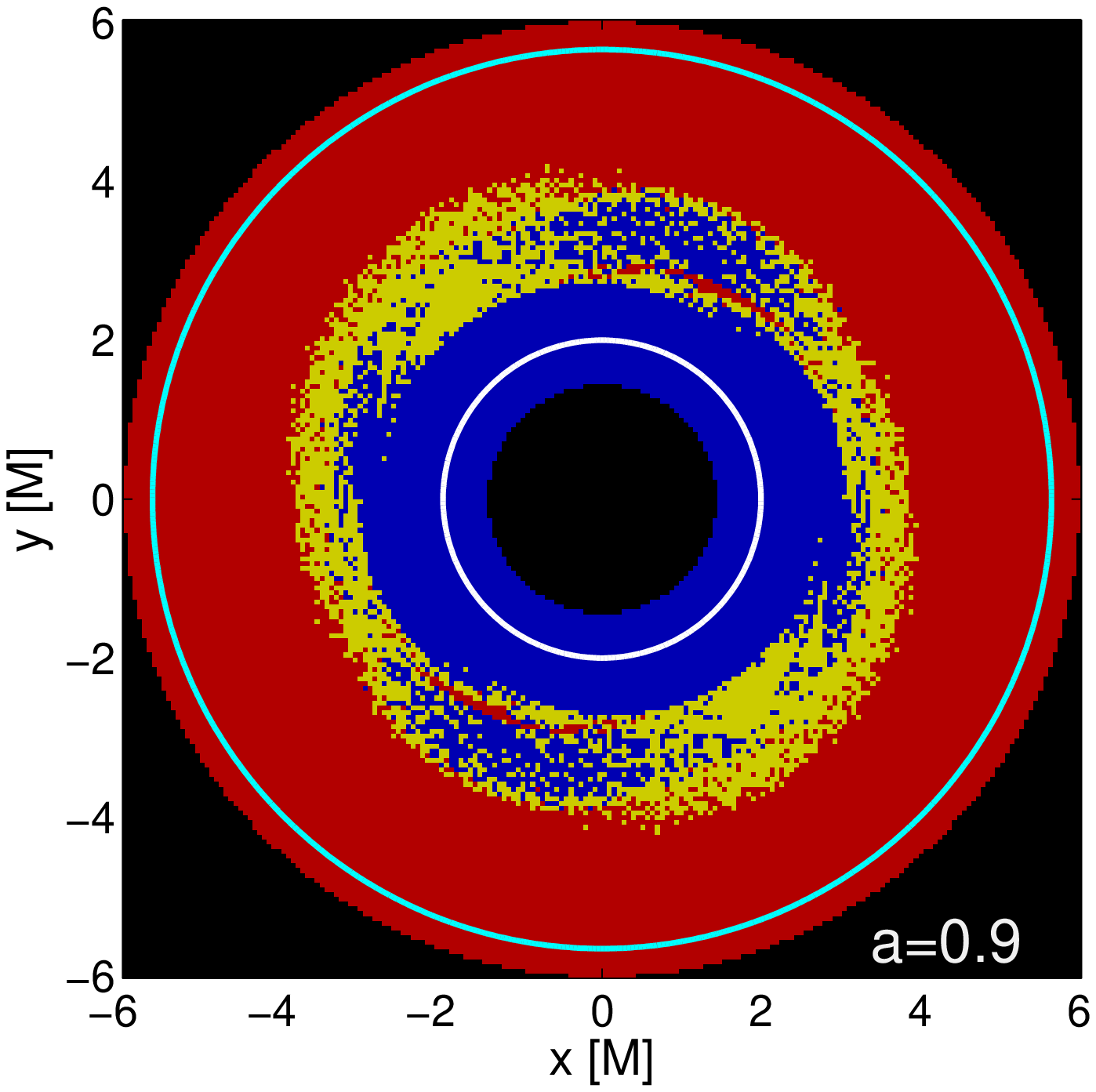}
\includegraphics[scale=.28]{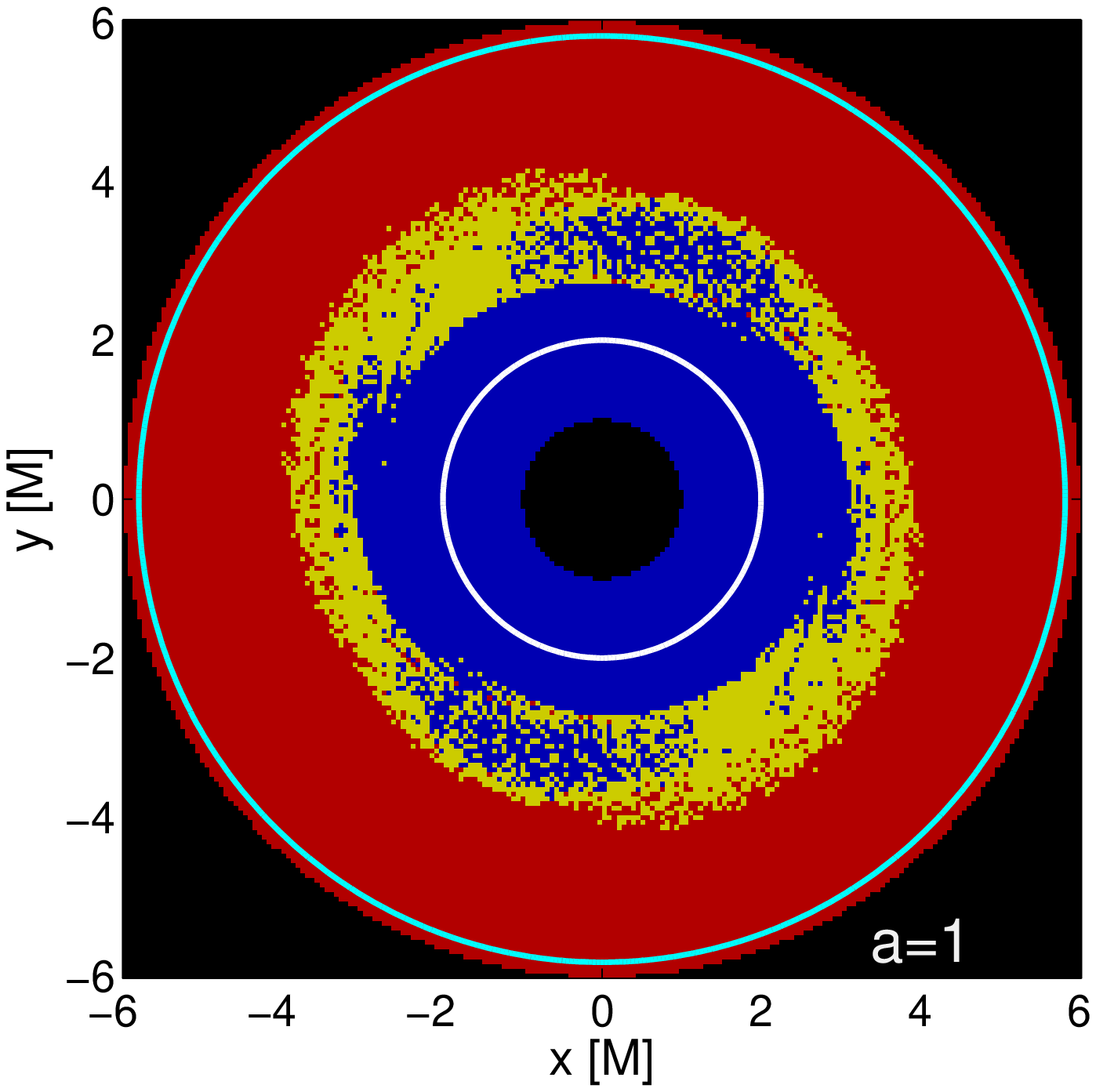}
\caption{Comparison of initially co-rotating orbits (upper row) with counter-rotating orbits (lower row). The color-coding and parameters as in Fig.~\ref{azi1}.}
\label{azi3}
\end{figure}

Analogously to our previous studies, we adopt the visual representation of resulting orbits in {escape boundary} plots in which we assign blue dots to plunging orbits, red dots to stable orbits and yellow dots to escaping ones. In Figs.~\ref{azi1}-\ref{azi3} we compare the  co-rotating setup (upper rows) with the counter-rotating case (bottom rows) in a series of escape boundary plots for a particular set of parameters ($qB=-5\,M^{-1}$ and $\alpha=35\degree$) with varying values of spin. In Fig.~\ref{azi1} we observe that for low spins the both cases do not differ significantly. Corresponding escape zones have comparable shapes and sizes regardless the sense of revolution. However, if we consider moderate values of spin, the situation changes and significant differences between the escape zones in co-rotating and counter-rotating disks arise as shown in Fig.~\ref{azi2}. Most importantly, we observe that the number of escaping trajectories is substantially higher for counter-rotating orbits forming considerably larger escape zones. The tendency of more rapid rotation to favorize the escape from counter-rotating orbits is confirmed in Fig.~\ref{azi3}, where we notice that high spin allows only narrow escape zones in the co-rotating disk while counter-rotating zones become significantly wider. Also, these plots exhibit clear signatures of a transition from the regular to chaotic system in agreement with previous studies of similarly perturbed black hole systems.\cite{1992GReGr..24..729K,1997CQGra..14.1249L,2015MNRAS.451.1770W,kopacek20} 

\begin{figure}[ht]
\center
\includegraphics[scale=.45]{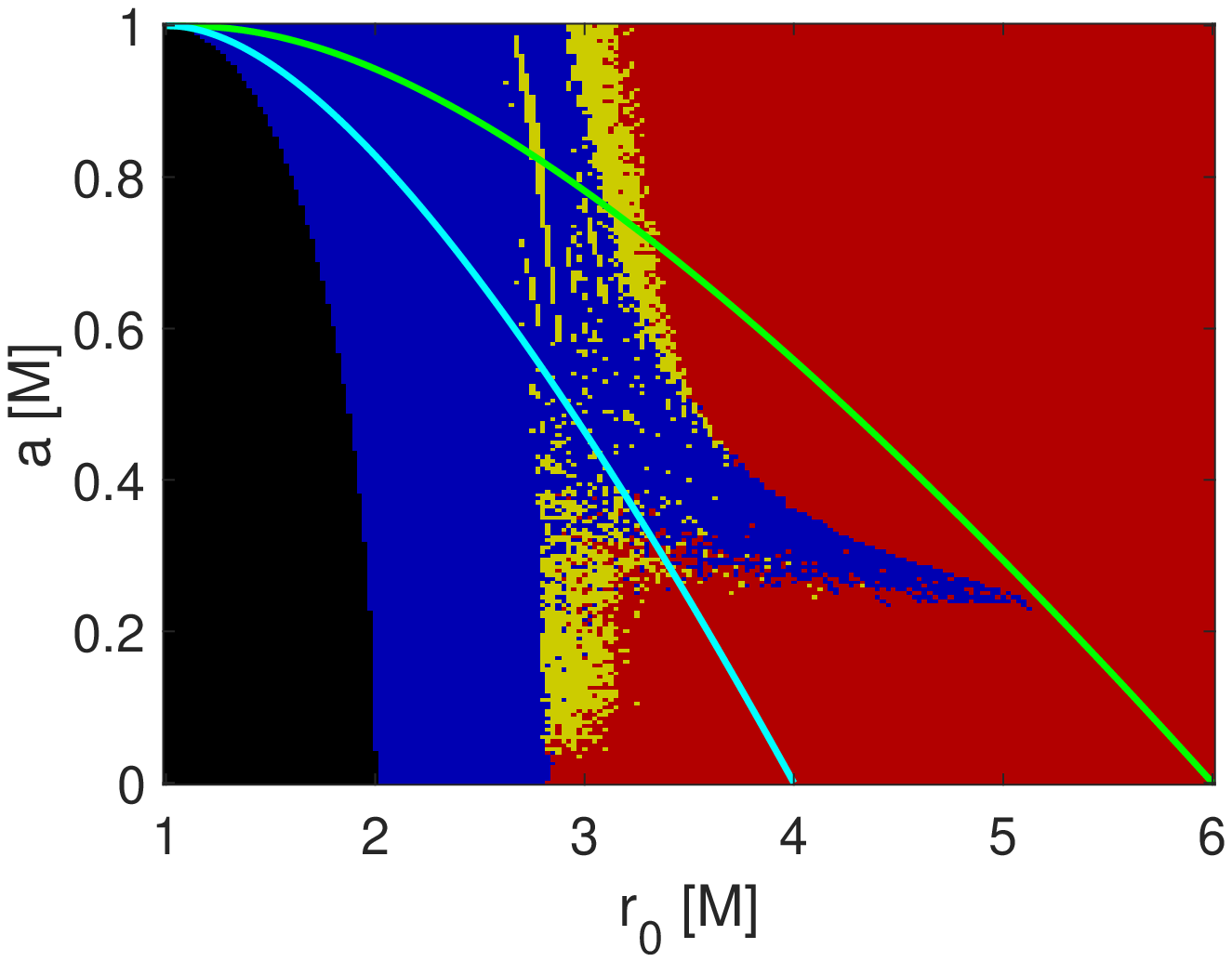}
\includegraphics[scale=.45]{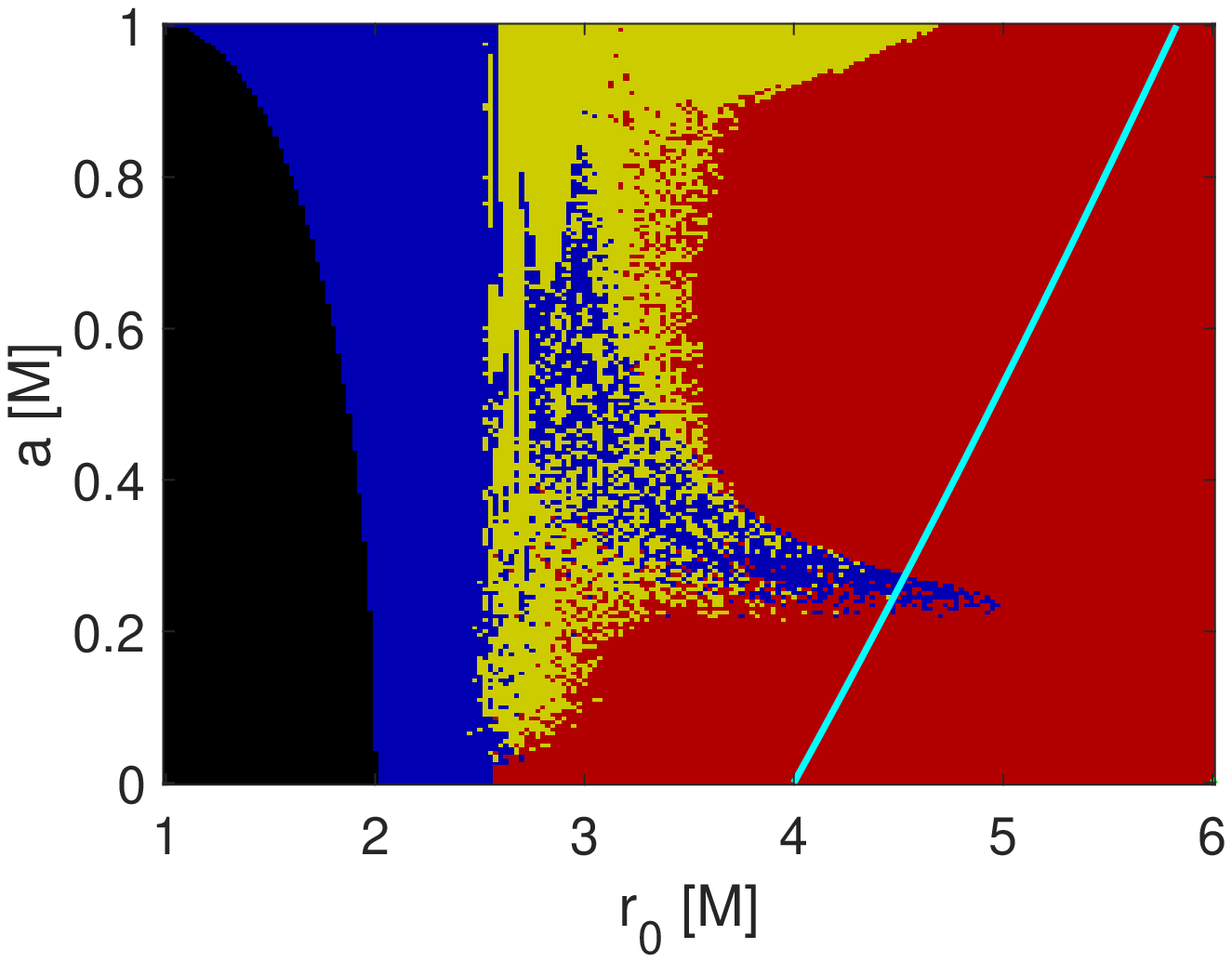}\\
\includegraphics[scale=.33]{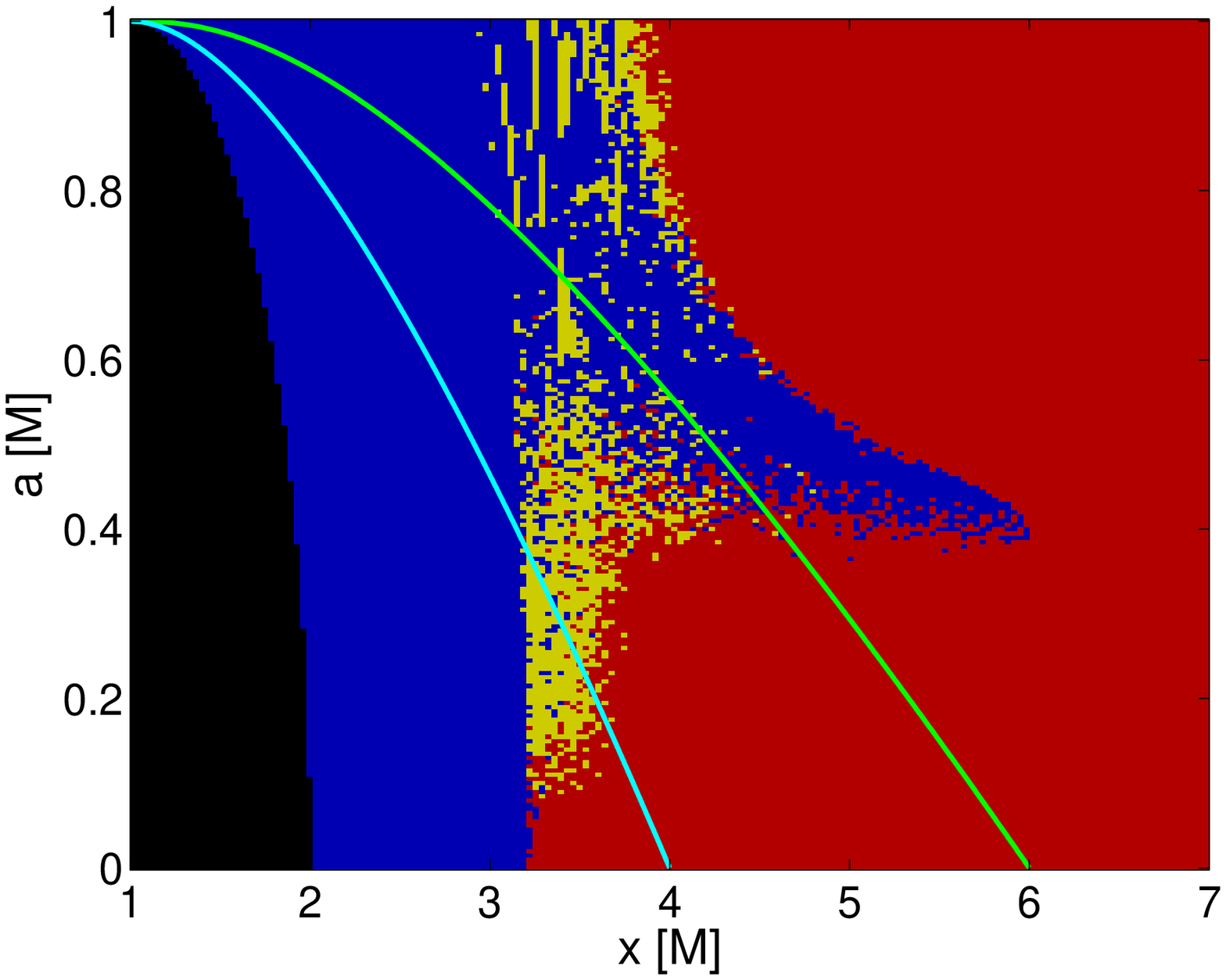}
\includegraphics[scale=.45]{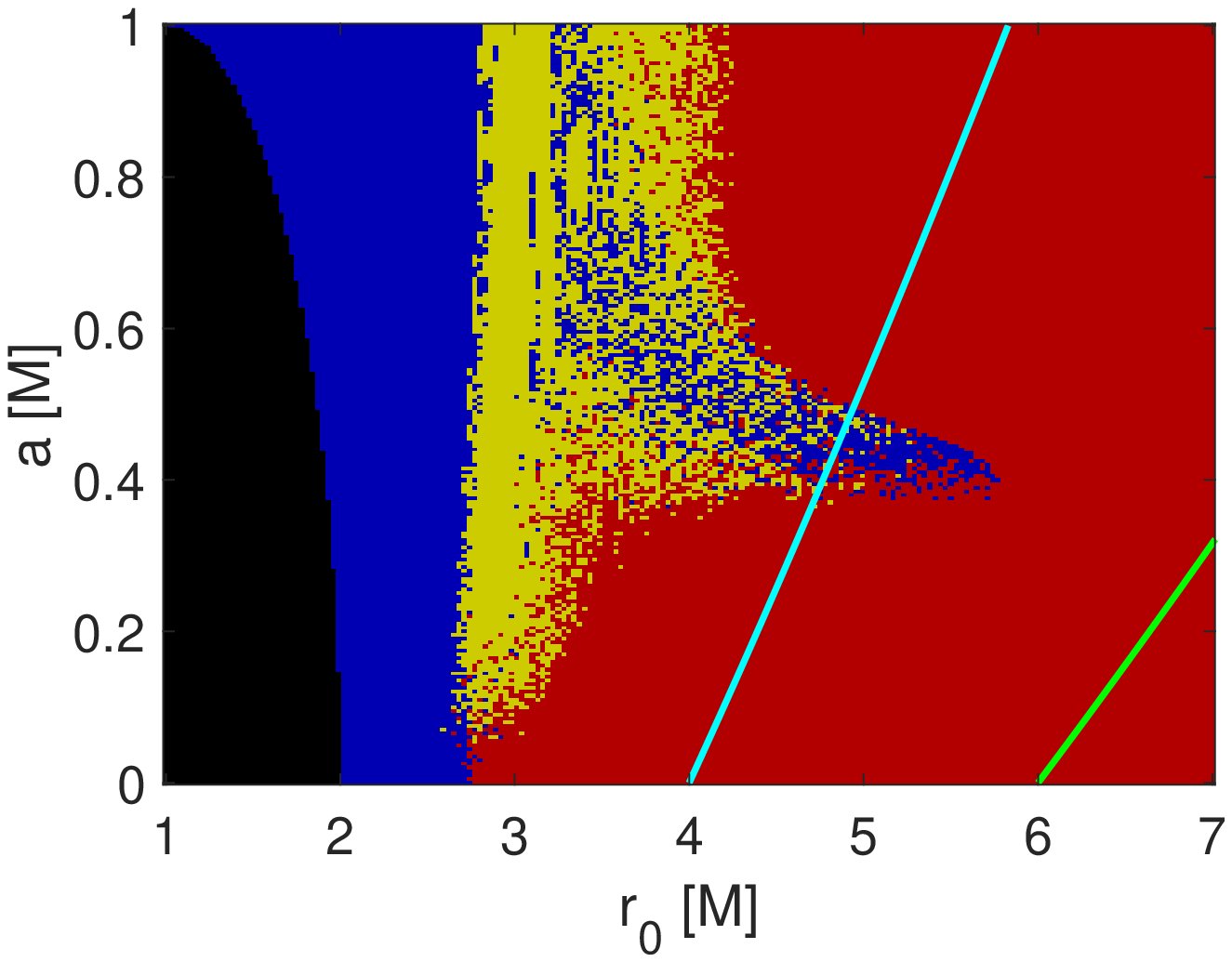}
\\
\includegraphics[scale=.45]{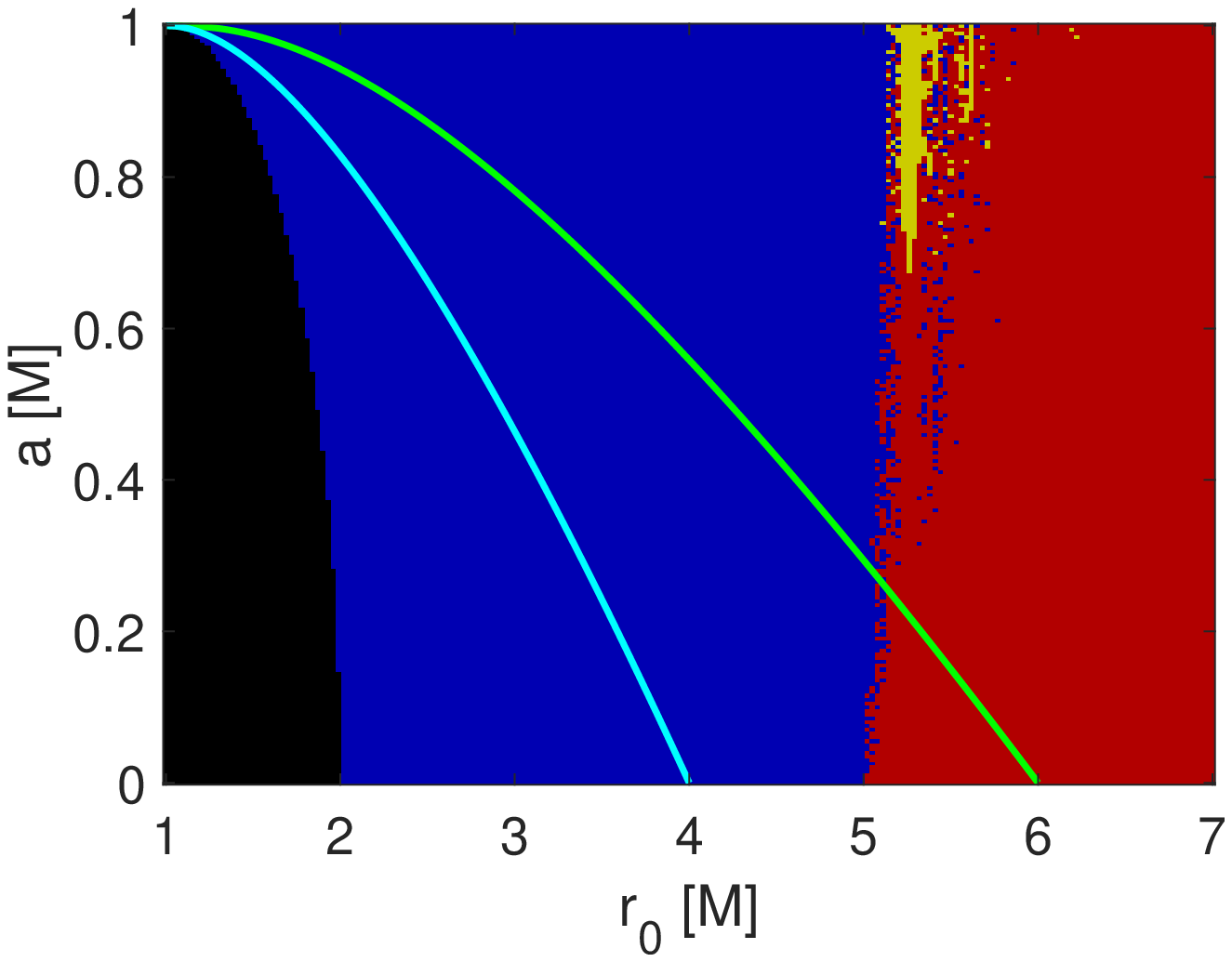}
\includegraphics[scale=.45]{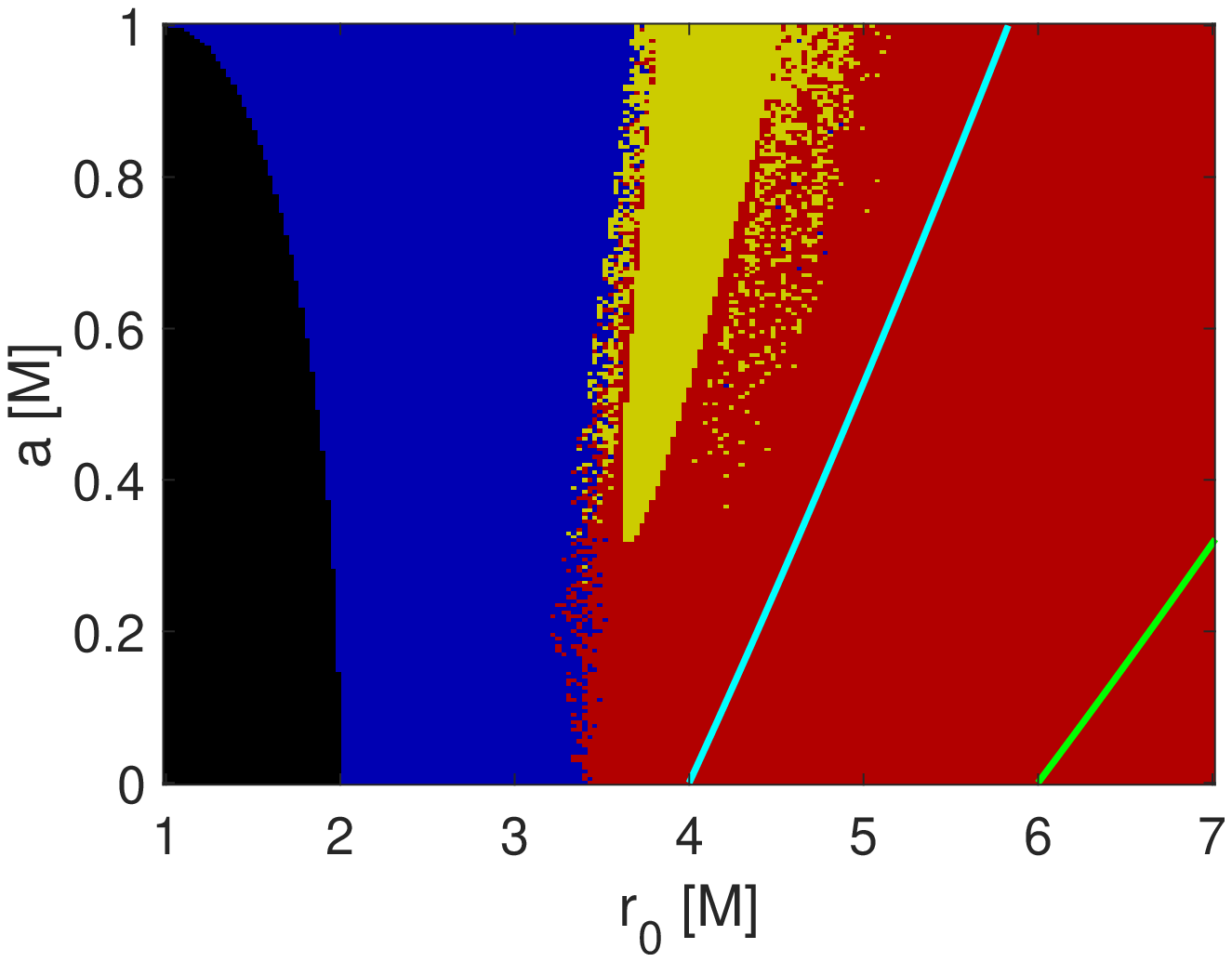}

\caption{Comparison of escape zones in co-rotating (left panels) and counter-rotating disks (right panels) for several values of magnetization parameter: $qB=-M^{-1}$ (bottom panels), $qB=-5\,M^{-1}$ (middle panels) and $qB=-10\,M^{-1}$ (upper panels). The inclination of the field is $\alpha=35\degree$ and $\varphi_0=135\degree$. The color-coding as in Fig.~\ref{azi1}.}
\label{escape1}
\end{figure}

Another convenient way to visualize the escape zones is to fix the initial value of the azimuthal angle $\varphi_0$ and plot the trajectories in the $r_0\times a$ plane (ionization radius vs. spin) using the same color-coding of trajectories as in Figs.~\ref{azi1}-\ref{azi3}. The angle $\varphi$ is measured anticlockwise from the positive direction of $x$-axis (which is the direction of the field inclination). In particular, we choose the value $\varphi_0=135\degree$ for which the structure of the escape zone seems especially complex. In \rff{escape1} we present the resulting plots for several values of magnetization parameter $qB$ and compare co-rotating escape zones (left column) with the counter-rotating zones (right column). It confirms that counter-rotating orbits are generally more prone to escape regardless the value of $qB$ while the both cases differ most significantly for rapidly rotating black holes. 

\begin{figure}[ht]
\center
\includegraphics[scale=.34, trim=12mm 0mm 0mm 0mm]{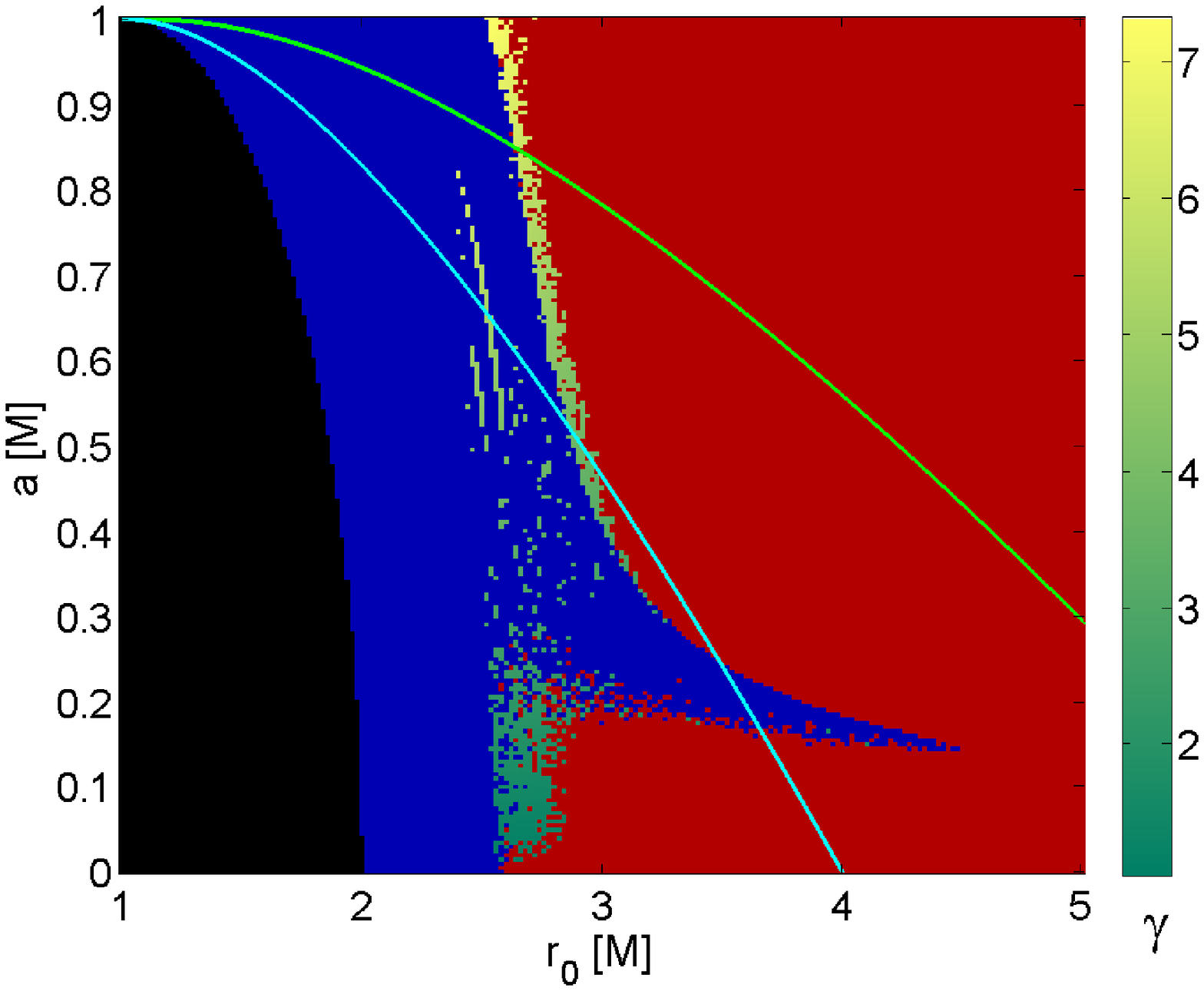}
\includegraphics[scale=.34, trim=10mm 0mm 0mm 0mm]{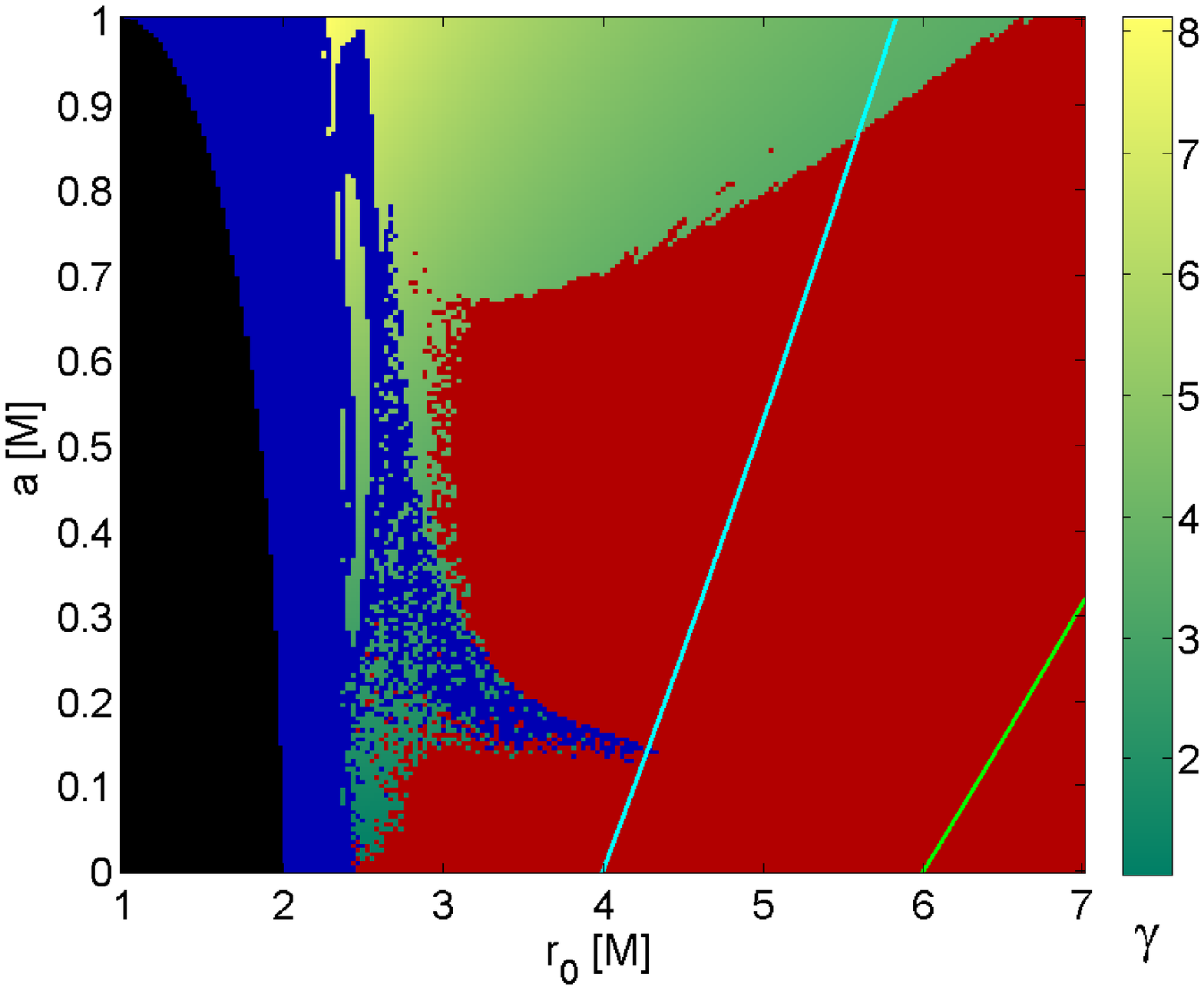}
\caption{Final Lorentz factor $\gamma$ of escaping particles is encoded with the colorscale comparing the co-rotating (left) and the counter-rotating disks (right) while other parameters remain fixed as: $\alpha=35\degree$ and $qB=-20\,M^{-1}$.}
\label{gamma}
\end{figure}

Previous analysis in Papers I and II shows that spin parameter is an important factor for the acceleration of escaping particles. In particular, it appears that final value of $\gamma$ for the fixed $qB$ grows with the spin and decreases with the ionization radius $r_0$. Maximally accelerated particles are thus expected to originate from the inner edge of the escape zone near maximally rotating black hole with spin $a=M$. 

In \rff{escape1} we observe that counter-rotating escape zones for high spins are considerably wider compared to co-rotating ones and, moreover, we notice that counter-rotating zones are located closer to the horizon. We thus expect that particles escaping from counter-rotating orbits could reach slightly higher energies. This is indeed the case as shown in \rff{gamma} where the final values of the Lorentz factor are encoded with the color-scale. For this particular set of parameters ($qB=-20\,M^{-1}$, $\varphi_0=135\degree$) we find that maximally accelerated counter-rotating particle reaches $\gamma_{\rm max}=8.1$ being launched from $r_0=2.3\,M$ while for the co-rotating disk we find $\gamma_{\rm max}=7.3$ for the particle escaping from $r_0=2.5\,M$.

\section{Conclusions}\label{concl}
We have numerically analyzed a simplified model of the outflow of electrically charged particles from the inner region of the accretion disk near a weakly magnetized rotating black hole. Stable circular orbits (freely falling below ISCO) of initially neutral particles were perturbed by the ionization process occuring in the non-axisymmetric magnetosphere. As a result, the escaping jet-like trajectories may be realized for some range of parameters and acceleration to high energies becomes possible. While we have previously studied various aspects of given model and investigated the formation of escape zones in detail,\cite{kopacek20,kopacek18} we have assumed the initial setup of co-rotating accretion disk. In particular, the neutral particles above ISCO were supposed to follow prograde circular orbits and freely fall below ISCO. 

In this contribution we have generalized the previous analysis by considering also counter-rotating (retrograde) Keplerian orbits as an initial setup of neutral matter. We have studied the escape zones emerging in the counter-rotating disks which were directly compared with the corresponding co-rotating versions. Based on the analyzed set of escape zones we may conclude that counter-rotation tends to increase the efficiency of the outflow. While its effect becomes negligible for low spins, for moderate to high spin values we observe that counter-rotating disks allow the formation of significantly wider escape zones compared to their co-rotating analogues. The position of the zone is also affected as counter-rotating zones are shifted towards the horizon which makes the acceleration process more efficient. Although the difference is rather small, we observe that counter-rotating particles may be accelerated to higher energies. 

\section*{Acknowledgements}
Authors acknowledge the support from the Inter-Excellence mobility program in science of the Czech Ministry of Education, Youth and Sports (projects ref.\ 8JCH 1080 and LTC 18058). OK has been supported by the project ``Lumina Quaeruntur'' (ref.\ LQ100032102) of the Czech Academy of Sciences. VK thanks the Czech Science Foundation grant ``Mass and charge currents in general relativity and astrophysics'' (ref.\ 21-11268S).

\bibliographystyle{ws-procs961x669}
\bibliography{article}

\end{document}